\documentclass{article}
\usepackage{apacite}\newcommand{\url}[1]{}
\usepackage{times}
\usepackage[dvips]{graphicx}
\usepackage{bm,amsfonts,amssymb,amsmath}
\usepackage{graphicx}
\let\mode=1
\newcommand{\deq}{\stackrel {\rm def}{=}}
\newcommand{\tom}{\tilde{\omega}}

\begin{document}

\date{}

\title{Introducing numerical bounds \\ to improve event-based neural network simulation}

\author{Bruno Cessac, Olivier Rochel, Thierry Vi\'eville \\ 
	LJAD {\footnotesize \tt \em http://math1.unice.fr} \& INRIA Cortex \& Odyssee {\footnotesize \tt \em http://inria.fr}}

\maketitle

\begin{abstract}

Although the spike-trains in neural networks are mainly constrained by the neural dynamics itself, 
global temporal constraints (refractoriness, time precision, propagation delays, ..) are also to be taken into account.
These constraints are revisited in this paper in order to use them in event-based simulation paradigms.

We first review these constraints, and discuss their consequences at the simulation level,
showing how event-based simulation of time-constrained networks can be simplified in this context:
the underlying data-structures are strongly simplified, while event-based and clock-based mechanisms can be easily mixed.
These ideas are applied to punctual conductance-based generalized integrate-and-fire neural networks simulation, while
spike-response model simulations are also revisited within this framework. 

As an outcome, a fast minimal complementary alternative with respect to existing simulation event-based methods,
with the possibility to simulate interesting neuron models is implemented and experimented.

\end{abstract}

\noindent {\bf\normalsize Key Words} Spiking network. Neural code. Simulation.

\section{Introduction}

Let us consider the simulation of large-scale networks of neurons,
in a context where the spiking nature of neurons activity is made explicit \cite{gerstner-kistler:02b},
either from a biological point of view or for computer simulation. 
From the detailed Hodgkin-Huxley model \cite{hodgkin-huxley:52}, 
(still considered as the reference but unfortunately intractable when considering neural maps), 
back to the simplest integrated and fire (IF) model, a large family of continuous-time
models have been produced, often compared with respect to their (i) biological plausibility and their (ii) simulation efficiency.

 As far as this contribution is concerned, we consider a weaker notion of biological plausibility: a simulation is biologically plausible if it verifies an explicit
set of constraints observed in biology. More precisely, we are going to consider a few global time constraints and develop their consequences at the 
simulation level.
It appears here that these biological temporal limits are very precious quantitative elements, allowing us on one hand 
to bound and estimate the coding capability of such systems and, on the other hand, to improve simulations.

\subsubsection*{Simulation efficiency of neural network simulators}

Simulation efficiency is a twofold issue of precision and performances. See \cite{brette-rudolph-etal:07} for a recent review about both event-based and clock-based
simulation methods.

Regarding precision, event-based simulations, in which firing times are not regularly discretized but calculated event by event at the machine precision level, 
provides (in principle) an {\em unbiased} solution.
On the reverse, it has been shown that a regular clock-based discretization of continuous neural systems may introduce systematic errors, 
with drastic consequences at the numerical level, even when considering very small sampling times \cite{rudolph-destexhe:07}.

Furthermore, the computational cost is in theory an order of magnitude better using event-based sampling methods \cite{brette:06}, 
although this may be not always the case in practice \cite{morrison-mehring-etal:05}, as further discussed in this paper.

However, using event-based simulation methods is quite tiresome: Models can be simulated if and only if the next spike-time can be explicitly computed in reasonable time.
This is the case only for a subset of existing neuron models
so that not all models can be used. 
An event-based simulation kernel is more complicated to use than a clock-based.
Existing simulators are essentially clock-based. Some of them integrate event-based as a marginal tool or 
in mixtures with clock-based methods \cite{brette-rudolph-etal:07}.
According to this collective review,
only the fully supported, scientifically validated, pure event-based simulators is MVASpike \cite{rochel-martinez:03},
the NEURON software proposing a well-defined event-based mechanism \cite{hine-carnevale:04}, 
while several other implementations (e.g.: DAMNED \cite{mouraud-puzenat-etal:06}, MONSTER \cite{rudolph-destexhe:06}) exists but are not publicly supported.

In other words, event-based simulation methods may save precision and computation time, but not the scientist time.

The goal of this paper is to propose solutions to overcome these difficulties, in order to have an easy to use unbiased simulation method.
 
\subsubsection*{Considering integrate and fire models.}

At the present state of the art, considering adaptive, bi-dimensional, non-linear, integrate-and-fire model with conductance based synaptic interaction
(as e.g. in  \cite{destexhe:97,brette-gerstner:05,rudolph-destexhe:07}), called ``punctual conductance based generalized integrate and fire models'' (gIF), 
presents several advantages: 

- They seem to provide an effective description of the neuronal activity allowing to reproduce several important neuronal regimes \cite{izhikevich:04}, 
with a good adequacy with respect to biological data, especially in high-conductance states, 
typical of cortical in-vivo activity \cite{destexhe-rudolph-etal:03}. 

- They provide nevertheless a simplification of Hodgkin-Huxley models, useful both for a mathematical analysis and numerical 
simulations \cite{gerstner-kistler:02b,izhikevich:03}.

In addition, though these models have mainly considered one neuron dynamics, they are easy to extend to network structure,
including synaptic plasticity \cite{markram-etal:97,pfister-gerstner:06}.
See, e.g. \cite{rauch-etal:03} for further elements in the context of experimental frameworks and \cite{la-camera-etal:08a,la-camera-etal:08b} for a review.

After a spike, it is assumed in such integrate and fired models that an {\em instantaneous} reset of the membrane potential occurs.
This is the case for all models except the Spike Response Model of \cite{gerstner-kistler:02b}.
From the information theoretic point of view, it is a temptation to relate this spurious property to the {\em erroneous} fact 
that the neuronal network information is not bounded.
In the biological reality, time synchronization is indeed not instantaneous (action potential time-course, synaptic delays, refractoriness, ..). 
More than that, these biological temporal limits are very precious quantitative elements, allowing one to bound and estimate the coding capability of the system. 

\subsubsection*{Taking time-constraints into account}

The output of a spiking neural network is a set of events, defined by their occurrence times, up to some precision:
\\ \centerline{${\cal F} = \{ \cdots t_i^n \cdots \}$, $t_i^1 < t_i^2 < \cdots < t_i^n < \cdots, \forall i, n$} \\
where $t_i^n$ corresponds to the $n$th spike time of the neuron of index $i$, with related inter-spike intervals $d_i^{n} = t_i^{n} - t_i^{n-1}$. 

In computational or biological contexts, not all sequences ${\cal F}$ correspond to spike trains since they are constrained by the neural dynamic, 
while temporal constraints are also to be taken into account \cite{cessac-etal:08a}.
This is the key point here: Spike-times are
\\~[C1] bounded by a refractory period $r$,
\\~[C2] defined up to some absolute precision $\delta t$, while 
\\~[C3] there is always a minimal delay $dt$ for one spike to propagate to another unit, and there might be (depending on the model assumptions) a 
\\~[C4] maximal inter-spike interval $D$ such that either 
the neuron fires within a time delay $< D$ or remains quiescent forever).
For biological neurons, orders of magnitude are typically, in milliseconds: 
\\ \centerline{\scriptsize \begin{tabular}{|c|c|c|c|} \hline $r$ & $\delta t$ & $d t$ & $D$ \\ \hline $1$ & $0.1$ & $10^{-[1, 2]}$ & $10^{[3, 4]}$ \\ \hline \end{tabular}}

The derivations of these numerical values are reviewed elsewhere \cite{cessac-etal:08a}. This has several consequence. 
On one hand, this allows us ti derive an upper bound for the amount of information:
\\ \centerline{$N \, \frac{T}{r} \, \log_2\left(\frac{T}{\delta t}\right)$ bits during $T$ seconds,}
while taking the numerical values into account it means for large $T$, a straight-forward numerical derivation leads to about $1 Kbits/neuron$.

On the other hand \cite{cessac:08,cessac-vieville:08}, it appears that for generalized integrate and fire neuron models with conductance synapses and constant external currents,
the raster plot is generically periodic, with arbitrary large periods, while there is a one-to-one correspondence between orbits and rasters. This last fact, and the fact 
that more general models such as Hodgkin-Huxley neuron assemblies can be simulated during a bounded period of time \cite{cessac-etal:08a}
provides theoretical justifications for the present work.

\subsubsection*{What is the paper about}

We develop here the consequences of the reviewed time constraints at the simulation level. 
Section~\ref{lazy-evaluation} shows how event-based simulation of time-constrained networks can be impacted and somehow improved in this context.
Section~\ref{gif-simul} considers punctual conductance based generalized integrate and fire neural network simulation, while
section~\ref{srm-simul} revisits spike-response model simulation within this framework. 
These mechanisms are experimented in section~\ref{results}, where the computer implementation choices are discussed.

Since the content of this paper requires the integration of data from the literature reused here,
we have collected these elements in the appendix in order to ease the main text reading, while maintaining the self-completeness of the contribution.

\section{Event-based simulation of time-constrained networks.}\label{lazy-evaluation}

Clock-based and event-based simulations of neural networks make already use at different level of the global time-constraints reviewed here.
See e.g. \cite{brette-rudolph-etal:07} for an introduction and a large review and \cite{rochel-martinez:03,brette:06,rudolph-destexhe:06,morrison-straube-etal:07} for 
simulations with event-based or hybrid mechanisms. However, it appears that existing event-based simulation mechanisms gain at being revisited.

In event-based simulation the exact simulation of networks of units (e.g. neurons) and firing events (e.g. spikes) fits in the discrete event system framework 
\cite{rochel-martinez:03} and is defined at the neural unit level by : \begin{quotation}
  -1- the calculation of the next event-time (spike firing),

  -2- the update of the unit when a new event occurs.
\end{quotation}

At the network level, the following two-stage mechanism completely implements the simulation:\begin{quotation}

  -a- retrieve the next event-time and the related unit,

  -b- require the update of the state of this unit, inform efferent units that this unit has emitted an event, and update the related event-times,

\end{quotation} repeating -a- and -b- when ever events occur or when some bound is reached.

This mechanism may also take external events into account (i.e., not produced by the network units, but by external mechanisms).

Such a strategy is thus based on two key features: \begin{itemize}

 \item The calculation of the next event-time, conditioned to the present state and to the fact that, by definition, no event is received in the meanwhile for each unit;

 \item The ``future'' event-time list, often named ``priority queue'', where times are sorted and in which event-times are retrieved and updated.

\end{itemize}

The goal of this section is to revisit these two features considering [C3] and [C4]. 

Let us consider in this section a network with $N$ units, an average of $C$ connections per units, with an average number $M \leq N$ of firing units.

\subsection{Event-time queue for time-constrained networks}

Although, in the general case, spike-times must be sorted among pending events, yielding a $O(\log(M))$ complexity for each insertion, 
there exists data structure allowing to perform retrieve/update operations in constant ``$O(1)$'' time. 
Several efficient data structures and algorithms have been proposed to handle such event scheduling task. They are usually based on heap-like structures \cite{rochel-martinez:03} or sets of buffers associated with some time intervals (such as the calendar queue in \cite{brette:06}). 
Ring buffers with fixed time step are used in \cite{morrison-mehring-etal:05}.

The basic idea of these structures is to introduce buckets containing event-times in a given time interval. 
Indexing these buckets allows one to access to the related times without considering what is outside the given time interval. 
However, depending on the fixed or adaptive bucket time intervals, bucket indexing mechanisms and times list managements inside a bucket, 
retrieve/update performances can highly vary.

Let us now consider [C3] and assume that the bucket time-interval is lower than $dt$, the propagation delay, 
lower than the refractory period $r$ and the spike time precision $\delta t$. If an event in this bucket occurs, there is at least a $dt$ delay before
influencing any other event. Since other events in this bucket occurs in a $dt$ interval they are going to occur before being influenced by another event.
As a consequence, they do not influence each others. They thus can be treated independently. 
This means that, in such a bucket, events can be taken into account in any order 
(assuming that for a given neuron the synaptic effect of incoming spikes can be treated in any order within a $dt$ window, 
since they are considered as synchronous at this time-scale).

It thus appears a simple efficient solution to consider a ``time histogram'' with $dt$ large buckets, as used in \cite{morrison-mehring-etal:05} under the name of ``ring buffer''.
This optimization is also available in \cite{rochel-martinez:03} as an option, 
while \cite{brette:06} uses a standard calendar queue, thus more general, but a-priory less tuned to such simulation.
Several simulation methods take into account [C3] (e.g. \cite{lee-etal:01,connolly-etal:04}).

The drawback of this idea could be that the buffer size might be huge. 
Let us now consider [C4], i.e. the fact that relative event-time are either infinite (thus not stored in the time queue) or bounded. 
In this case with $D = 10^{[3, 4]}ms$ (considering fire rate down to $0.1 Hz$) and $10^{-[1, 2]}$ (considering gap junctions) the buffer size $S = D/dt = 10^{5-6}$,
which is easily affordable with computer memories. 
In other words, {\em thanks to biological order of magnitudes reviewed previously, the histogram mechanism appears to be feasible}.

If [C4] does not hold, the data-structure can be easily adapted using a two scale mechanism. 
A value of $D$, such that almost all relative event-time are lower than $D$ is to be chosen.
Almost all event-times are stored in the initial data structure, whereas larger event-times are stored in a transient calendar queue before being reintroduced 
in the initial data structure. This add-on allows one to easily get rid of [C4] if needed, and still makes profit of the initial data structure for almost all events.
In other words, this idea corresponds to considering a sliding window of width $D$ to manage efficiently events in a near future.
This is not implemented here, since models considered in the sequel verify [C4].

The fact that we use such a time-histogram and treat the events in a bucket in any order allows us to drastically simplify and speed-up the simulation. 

Considering the software implementation evaluated in the experimental section, we have observed the following.
Event retrieval requires less than 5 machine operations and event update less than 10, including the on-line detection of [C3] or [C4] violation.
The simulation kernel\footnote{\label{enas} Source code available at {\tt http://enas.gforge.inria.fr}.} is minimal (a 10Kb C++ source code), 
using a ${\cal O}(D/dt+N)$ buffer size and about ${\cal O}(1+C+\epsilon/dt) \simeq 10-50$ operations/spike, 
thus we a small overhead $\epsilon \ll 1$, corresponding to the histogram scan.
In other words, the mechanism is ``$O(1)$'' as for others simulation methods. Moreover, the time constant is minimal in this case.
Here, we save computation time, paying the price in memory.

\subsubsection*{Remarks}~\vspace{-0.5cm}

The fact that we use such a time-histogram does not mean that we have discretized the event-times. The approximation only concerns the way how events are processed. 
Each event time is stored and retrieved with its full numerical precision. Although [C2] limits the validity of this numerical value, 
it is indeed important to avoid any additional cumulative round-off. 
This is crucial in particular to avoid artificial synchrony \cite{rochel-martinez:03,morrison-mehring-etal:05}.

Using [C3] is not only a ``trick''. It changes the kind of network dynamics that can be simulated. 
For instance, consider a very standard integrate and fire neuron model. 
It can not be simulated in such a network, since it can instantaneously fire after receiving a spike,
whereas in this framework adding an additional delay is required.
Furthermore, avalanche phenomena (the fact that neurons instantaneously fire after receiving a spike, instantaneously driven other neurons and so on..)
cannot occur.
A step further, temporal paradoxes (the fact, e.g., that a inhibitory neuron instantaneously fires inhibiting itself thus . . is not supposed to fire)
cannot occur and have not to be taken into account. When considering the simulation of biological systems, [C3] indeed holds.

Only sequential implementation is discussed here. The present data structure is intrinsically sequential. 
In parallel implementations, a central time-histogram can distribute the unit next-time and state update calculation on several processors,
with the drawback that the slower calculation limits the performance.
Another idea is to consider several time-histograms on different processors and synchronize them. 
See \cite{mouraud-puzenat-etal:06} and \cite{morrison-straube-etal:07} for developments of these ideas.

The fact that we use such a tiny simulation kernel has several practical advantages. e.g. to use spiking network mechanisms in embedded systems, etc.. 
However, it is clear that this is not ``yet another'' simulator because a complete simulator requires much more that an event queue \cite{brette-rudolph-etal:07}. 
On the contrary, the implementation has been designed to be used as a plug-in in existing simulators, mainly MVASpike \cite{rochel-martinez:03}. 

\subsection{From next event-time to lower-bound estimation}

Let us now consider the following modification in the event-based simulation paradigm. Each unit provides: \begin{quotation}
  -1'- the calculation of {\em either} the next event-time, 
  \\ {\em or} the calculation of a \underline{lower-bound} of the next event-time,

  -2- the update of the neural unit when an internal or external event occurs, 
  \\ {\em with} the indication whether the previously provided next event-time or lower-bound is still valid.
\end{quotation}

At the network level the mechanism's loop is now:\begin{quotation}
 -a'- retrieve {\em either} the next event-time and proceed to -b'- {\em or} a lower-bound and proceed to -c-

 -b'- require the update of the state of this unit, inform efferent units that this unit has emitted an event, 
   and update the related event-times {\em only} if this event-time is lower than its previous estimation,

 -c- store the event-time lower-bound in order to re-ask the unit at that time.
\end{quotation}

A simple way to interpret this modification is to consider that a unit can generate ``silent events'' which write: ``Ask me again at that time, I will better know''.  

As soon as each unit is able to {\em provide the next event-time after a finite number of lower-bound estimations}, the previous process is valid.

This new paradigm is fully compatible with the original, in the sense that units simulated by the original mechanism are obviously simulated here, 
they simply never return lower-bounds.

It appears that the implementation of this ``variant'' in the simulation kernel is no more than a few additional line of codes. 
However, the specifications of an event-unit deeply change, since the underlying calculations can be totally different.

We refer to this modified paradigm as the {\em lazy} event-based simulation mechanism.

The reason of this change of paradigm is twofold: \begin{itemize}

\item Event-based and clock-based calculations can be easily mixed using the same mechanism.

Units that must be simulated with a clock-based mechanism simply return the next clock-tick as lower-bound, unless they are ready to fire an event.
However in this case, each unit can choose its own clock, requires low-rate update when its state is stable or require higher-rate update when in transient stage, etc..
Units with different strategies can be mixed, etc.. 

For instance, in \cite{morrison-mehring-etal:05} units corresponding to synapses are calculated in event-based mode, 
while units corresponding to the neuron body are calculated in clock-based mode, minimizing the overall computation load.
They however use a more complicated specific mechanism and introduce approximations on the next spike-time calculations.

At the applicative level, this changes the point of view with respect to the choice of event-based/clock-based simulations. 
Now, an event-based mechanism can always simulate clock-based mechanisms, using this useful trick.

\item Computation time can be saved by postponing some calculations.

Event-based calculation is considered as costless than clock-based calculation because the neuron state is not recalculated at each time-step, only when a new event is input.
However, as pointed out by several authors, when a large amount of events arrive to a unit, the next event-time is recalculated a large amount of time which can be much higher
than a reasonable clock rate, inducing a fall of performances.

Here this drawback can be limited. When a unit receives an event, it does not need to recalculate the next event-time, 
as soon as it is known as lower that the last provided event-time bound.
This means that if the input event is ``inhibitory'' (i.e. tends to increase the next event-time) or if the unit is not ``hyper-polarized'' 
(i.e. not close to the firing threshold, which is not trivial to determine) the calculation can be avoided, 
while the opportunity to update the unit state again later is to be required instead.

\end{itemize}

\subsubsection*{Remarks}~\vspace{-0.5cm}

Mixing event-based and clock-based calculations that way is reasonable, only because the event-time queue retrieve/update operations have a very low cost. 
Otherwise, clock-ticks would have generated prohibitory time overloads.

Changing the event-based paradigm is not a simple trick and may require to reconsider the simulation of neural units. 
This is addressed in the sequel for two important classes of biologically plausible neural units, at the edge of the state of the art and of common use:
Synaptic conductance based models \cite{destexhe:97} and spike response models \cite{gerstner-kistler:02b}.

\section{Event-based simulation of adaptive non-linear gIF models}\label{gif-simul}

Let us consider a {\em normalized} and {\em reduced} ``punctual conductance based generalized integrate and fire'' (gIF) neural unit model 
\cite{destexhe:97} as reviewed in \cite{rudolph-destexhe:06}.

The model is normalized in the sense that variables have been scaled and redundant constants reduced. This is a standard usual one-to-one transformation, 
discussed in the next subsection.

The model is reduced in the sense that both {\em adaptive} currents and non-linear {\em ionic} currents are no more explicitly depending on the potential membrane,
but on time and previous spikes only. This is a non-standard approximation and a choice of representation carefully discussed in appendix~\ref{gif-reduction}.

Let $v$ be the normalized membrane potential and $\tom_t = \{ \cdots t_i^n \cdots \}$ the list of all spike times $t_i^n < t$.
Here $t_i^n$ is the $n$-th spike-time of the neuron of index $i$.
The dynamic of the integrate regime writes: 
\begin{equation}\label{gIFi} \frac{dv}{dt} + g(t, \tom_t) \, v = i(t, \tom_t), \end{equation}
while the fire regime (spike emission) writes $v(t) = 1 \Rightarrow v(t^+) = 0$ with a firing threshold at 1 and a reset potential at 0, for a normalized potential.

Equation~(\ref{gIFi}) expands:
\begin{equation}\label{gIFe} \frac{dv}{dt} + \frac{1}{\tau_L} \left[v - E_L\right] +
 \sum_{j} \sum_{n} r_j\left(t - t_j^n\right) \left[v - E_j\right]
 = i_m(\tom_t), \end{equation}
where $\tau_L$ and $E_l$ are the membrane leak time-constant and reverse potential, 
while $r_j()$ and $E_j$ the spike responses and reverse potentials for excitatory/inhibitory synapses and gap-junctions
as made explicit in appendix~\ref{gif-normalization}.
Here, $i_m()$ is the reduced membrane current, including simplified adaptive and non-linear ionic current.

\subsection{Reduction of internal currents} \label{gif-reduction}

 Let us now discuss choices of modeling for $i_m = I^{adapt} + I^{ionic}$

\subsubsection*{Adaptive current}

In the Fitzhugh-Nagumo reduction of the original Hodgkin-Huxley model \cite{hodgkin-huxley:52} the average kinematics of the membrane channels
is simulated by a unique adaptive current. Its dynamics is thus defined, between two spikes, by a second equation of the form:
\[
 \tau_w \, \frac{dI^{adp}}{dt} = g_w \, \left( V - E_L \right) - I^{adp} + \Delta_w \, \delta\left(V - V_{\mbox{threshold}}\right), 
\]
with a slow time-constant adaptation $\tau_w \simeq 144 ms$, a sub-threshold equivalent conductance $g_w \simeq 4 nS$ and a level $\Delta_w \simeq 0.008 nA$ of spike-triggered
adaptation current. It has been shown \cite{izhikevich:03} that when a model with a quadratic non-linear response is increased by this adaptation current, it can be tuned to 
reproduce qualitatively all major classes of neuronal in-vitro electro-physiologically defined regimes.

Let us write:
\[
\begin{array}{rcl}I^{adp}(V,t) 
    &=& e^{\frac{-(t-t_0)}{\tau_w}} \, I^{adp}(t_0) + \frac{g_w}{\tau_w} \, \int_{t_0}^t e^{\frac{-(t-s)}{\tau_w}} (V(s) - E_L) ds + \Delta_w \, \#(t_0, t) \\
    &\simeq& \underbrace{e^{\frac{-(t-t_0)}{\tau_w}} \, I^{adp}(t_0) + g_w \,\left(1 - e^{\frac{-(t-t_0)}{\tau_w}}\right) \,(\bar{V} - E_L)}_{slow\;variation} + 
	\underbrace{\Delta_w \, \#(t_0, t)}_{spike-time\;dependent} \\
\end{array}
\]
where $\#(t_0, t)$ is the number of spikes in the $[t_0, t]$ interval while $\bar{V}$ is the average value between $t$ and $t_0$, the previous spike-time. 

Since the time-constant adaptation is slow, and since the past dependency in the exact membrane potential value is removed when resetting, 
the slow-variation term is almost constant.
This adaptive term is mainly governed by the spike-triggered adaptation current, the other part of the adaptive current being a standard leak.
This is also verified, by considering the linear part of the differential system of two equations in $V$ and $I^{adp}$, 
for an average value of the conductance $\bar{G}^+ \simeq 0.3\cdots1.5 nS$ and $\bar{G}^- \simeq 0.6\cdots2.5 nS$. 
It appears that the solutions are defined by two decreasing exponential profiles with $\tau_1 \simeq 16 ms << \tau_2 \simeq 115 ms$ time-constants, the former
being very close to the membrane leak time-constant and the latter inducing very slow variations.

In other words {\em current adaptation is, in this context, mainly due to spike occurrences} and 
the adaptive current is no more directly a function of the membrane potential but function of the spikes only.

\subsubsection*{Non-linear ionic currents}

Let us now consider the non-linear active (mainly Sodium and Potassium) currents responsible for the spike generation.
In models designed to simplify the complex structure of Hodgkin-Huxley equations, the sub-threshold membrane potential is defined by a supra-linear kinematics,
taken as e.g. quadratic or exponential, the latter form closer to observed biological data \cite{brette-gerstner:05}. It writes, for example:
\begin{equation} \label{Iexp}
I^{ion}(V) = \frac{C \, \delta_a}{\tau_L} e^{\frac{V-E_a}{\delta_a}} \geq 0 \mbox{ with } \left.\frac{d  I^{ion}}{d V}\right|_{V=E_a} = \frac{C}{\tau_L}
\end{equation}
with $E_a \simeq -40mV$ the threshold membrane state at which the slope of the I-V curve vanishes, while $\delta_a = 2 mV$ is the slope factor which determines the
sharpness of the threshold. There is no need to define a precise threshold in this case, since the neuron fires when the potential diverges to infinity.

A recent contribution \cite{touboul:08} re-analyzes such non-linear currents, 
proposes an original form of the ionic current, with an important sub-threshold characteristic not present in previous models \cite{izhikevich:03,brette-gerstner:05} 
and show that one obtains the correct dynamics, provided that the profile is mainly non-negative and strictly convex.
This is not necessarily a quadratic or exponential function.

Making profit of this general remark, we propose to use a profile of the form of~(\ref{Iexp}),
but simply freeze the value of $V$ to the the previous value obtained at the last spike time occurrence.
This allows us to consider a supra-linear profile which depends only on the previous spike times\footnote{In fact a more rigorous result can be derived,
although at the implementation level, the simple heuristic proposed here seems sufficient.
Let us write $i(V,t, \tom_t) =  i'(t, \tom_t) + I^{(ion)}(V, t, \tom_t)$ thus separate the $I^{(ion)}$ from all other currents written $i'(t, \tom_t)$.
Let us consider the last spike time $t_0$  of this neuron and let us write $\tilde{V}$ 
the solution of the linear differential equation ``without'' the ionic current $I^{(ion)}$:
\\ \centerline{$C \, \frac{d\tilde{V}}{dt}+ g(t,\tom_t) \, \tilde{V} = i'(t,\tom_t)$} \\
with $V(t_0) = V_{reset}$, as obtained above.
Define now $\hat{V} = V - \tilde{V}$, with $\hat{V}(t_0) = 0$, $V$ being the solution of the previous equation (without the ionic current). This yields:
\\ \centerline{$C \, \frac{d \hat{V}}{dt} + g(t,\tom_t) \, \hat{V} = I^{(ion)}(\hat{V} + \tilde{V}(t,\tom_t), t, \tom_t)$} \\
as easily obtained by superposition of the linear parts of the equation.
\\ Let $h(t, \tom_t)$ be any regular function and $f(V)$ any bijective regular function with $f(\hat{V}) \neq 0$. 
These two functions allow us to model a whole family of ionic currents:
\\ \centerline{$I^{(ion)}(\hat{V} + \tilde{V}(t,\tom_t), t,\tom_t) = g(t,\tom_t) \, \hat{V} + \frac{h(t,\tom_t)}{f(\hat{V})}.$}\\
The choice of $h$ and $f$ is simply related to specific properties: The reader can easily verify, by a simple integration, that it allows to obtain a closed form:
\\ \centerline{$\hat{V}(t,\tom_t) = F^{-1}\left(\int_{t_0}^t h(s,\tom_t) \, ds\right) \mbox{ with } F' = f \mbox{ and } F(0) = 0.$} \\
so that $\hat{V}$ is now a function of $\tom_t,t$ with $\hat{V}(t_0,\tom_t) = 0$, and so is $I^{(ion)}(\hat{V}(t,\tom_t) + \tilde{V}(t,\tom_t), t, \tom_t)$, 
removing the direct dependence on $V$. 
In other words, it now depends only on $t$ and on the spike times (thus on $\tom_t$) and not anymore on the membrane potential explicitly.
Clearly, this only applies to neurons which have fired at least once during the period of observation. Otherwise, we assume that its initial condition was also $V_{reset}$.
\\ We can, .e.g, choose:
\\ \centerline{$\begin{array}{rcl}
    I^{(ion)}(V, t,\tom_t)  &=& \frac{C \, \delta_a}{\tau_L} e^{\frac{V(t)-E_a(t,\tom_t)}{\delta_a}} \\
    E_a(t,\tom_t) &=& \tilde{V}(t,\tom_t) - \delta_a \, ln\left(\frac{g(t, \tom_t)}{\bar{g}}\right) \\
 \end{array}$} \\
for any $\bar{g} > 0$ which allows to control the threshold for different conductance. 
\\ Here $h = g$ and $f(v) = (k \, e^{\frac{v}{\delta_a}} - v)^{-1}$ for some $k$.
\\ In this case the threshold is no more fixed, but adaptive with respect to $g(t)$: the higher the conductance, the higher the threshold (via the $\bar{V}$). 
This is coherent with what has been observed experimentally \cite{azouz-gray:00,wilson-weyrick:04}, since the higher the conductance, 
the higher the frame rate increases with the spiking threshold.}. 
This approximation may slightly underestimate the ionic current before a spike, since $V$ is increasing with time. However, when many spikes are input, as it is the case
for cortical neurons, errors are minimized since the ionic current update is made at high rate.

At a phenomenological level \cite{izhikevich:03} the real goal of this non-linear current in synergy with adaptive currents is to provide several firing regimes. 
We are going to verify experimentally that even coarser approximations allow to attain this goal.

\subsection{Derivation of a spike-time lower-bound.}

Knowing the membrane potential at time $t_0$ and the list of spike times arrival, one can obtain the membrane potential at time $t$, from~(\ref{gIFi}):
\begin{equation} \label{IFuSol} v(t) = \nu(t_0, t, \tom_{t_0}) \, v(t_0) + \int_{t_0}^t \nu(s, t, \tom_{t_0}) \, i(s, \tom_{t_0}) \, ds \end{equation}
with:
\begin{equation} \label{IFuSolNu} \log(\nu(t_0, t_1, \tom_{t_0})) = -\int_{t_0}^{t_1} g(s, \tom_{t_0}) \, d s \end{equation}

Furthermore, 
\begin{equation}\label{gIFd} \begin{array}{rclcl}
g(t, \tom_{t_0}) &=& \frac{1}{\tau_L} + \sum_{j} \sum_{n} r_j\left(t - t_j^n\right) &>& 0 \\
i(t, \tom_{t_0}) &=& \frac{1}{\tau_L} \, E_L + \sum_{j} \sum_{n} r_j\left(t - t_j^n\right) \, E_j + i_m(\tom_{t_0}) &\ge& 0 \\
\end{array}\end{equation}
since the leak time-constant, the conductance spike responses are positive, the reverse potential are positive (i.e. they are larger than or equal to the reset potential) and 
the membrane current is chosen positive.

\begin{figure}[ht]  
 \begin{center} \includegraphics[width=12cm]{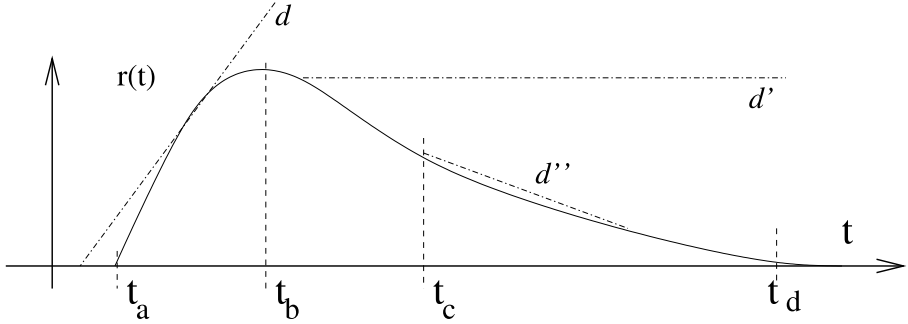} \end{center} 
 \caption{\small \label{response} The spike response profile. 
It has a flat response during the absolute delay interval $[0 ,  t_a]$, 
an increasing convex profile until reaching its maximum at $t_b$, 
followed by a decreasing convex and then concave profile, with an inflexion point at $t_c$. After $t_d$ the response is negligible. 
See text for details about $d$, $d'$ and $d''$.}
\end{figure}

The spike-response profile schematized in Fig.~\ref{response} and a few elementary algebra yields to the following bounds:
\begin{equation}\label{rBound} t \in [t_0 ,  t_1] \Rightarrow r_{\land}(t_0, t_1) \leq r(t) \leq r_{\lor}(t_0, t_1) + t \, r'_{\lor}(t_0 , t_1) \end{equation}
\\ writing $r'$ the time derivative of $r$, with 
\\ \centerline{$r_{\land}(t_0, t_1) = \min(r(t_0), r(t_1)) \mbox{ and } r_{\lor}(t_0, t_1) = \max(r(t_0), r(t_1))$} 
\\ with a similar definition for $r'_{\lor}(t_0 , t_1)$.

Here we thus consider a constant lower-bound $r_{\land}$ and a linear or constant upper-bound $r_{\lor} + t \, r'_{\lor}$. 
The related two parameters are obtained considering in sequence the following cases: 
\\ \vspace{0.2cm}\centerline{\scriptsize \begin{tabular}{c|c|c|c|c|} & $t_0 \in$  & $t_1 \in$ & $r_{\lor}(t_0, t_1)$ & $r'_{\lor}(t_0 , t_1)$ \\ \hline
(i)   & $[t_1 - r(t_1) / r'(t_1) ,  t_a]$ & $[t_a ,  t_b]$ & $r(t_1) - t_1 \, r'(t_1)$ & $r'(t_1)$ \\ \hline
(ii)  & $[t_a ,  t_b]$ & $[t_0 ,  +\infty]$ & $r(t_0) - t_0 \, r'(t_0)$ & $r'(t_0)$ \\ \hline
(iii) & $[t_c ,  +\infty]$ & $]t_0 ,  +\infty]$ & $(r(t_0) \, t_1 - r(t_1) \, t_0) / (t_1 - t_0)$ & $(r(t_1) - r(t_0)) / (t_1 - t_0)$ \\ \hline
(iv)  & $]-\infty , t_1]$ & $[t_0, t_b]$ & $r(t_1)$ & $0$  \\ \hline
(v)   & $]-\infty , t_b]$ & $[t_b ,  +\infty]$ & $r(t_b)$ & $0$  \\ \hline
(vi)  & $[t_b , t_1]$ & $[t_0 , +\infty]$ & $r(t_0)$ & $0$  \\ \hline
\end{tabular}}\vspace{0.2cm}

In words, conditions (i) and (ii) correspond to the fact that the the convex profile is below its tangent (schematized by $d$ in Fig.~\ref{response}),
condition (iii) that the profile is concave (schematized by $d''$ in Fig.~\ref{response}). 
In other cases, it can be observed that 1st order (i.e. linear) bounds are not possible.
We thus use constant bounds (schematized by $d'$ in Fig.~\ref{response}). 
Conditions (iv) and (vi) correspond to the fact the profile is monotonic, while condition (v) corresponds to the fact the profile is convex. 
Conditions (iv), (v) and (vi) correspond to all possible cases. Similarly, the constant lower bound corresponds to the fact the profile is either monotonic or convex.

From these bounds we derive: \begin{equation} \label{gIGlb1} \begin{array}{rcl}
 g(t, \tom_{t_0}) &\geq& \frac{1}{\tau_{\land}(t_0, t_1)} \\ &\deq& \frac{1}{\tau_L} + \sum_{j} \sum_{n} r_{j\land}(t_0, t_1) \\
 i(t, \tom_{t_0}) &\leq& i_{\lor}(t_0, t_1) + t \ i'_{\lor}(t_0, t_1) \\ &\deq& 
   \left [\frac{1}{\tau_L} \, E_L + \sum_{j} \sum_{n} r_{j\lor}(t_0, t_1)  \, E_j + i_m(\tom_{t_0}) \right] \\ &+& t \,
   \left [\sum_{j} \sum_{n} r'_{j\lor}(t_0, t_1) \, E_j \right] \\
\end{array} \end{equation}
while $i'_{\lor}(t_0, t_1) \geq 0$ as the positive sum of $r'_{j\lor}$ values is always positive in our case.

Combining with~(\ref{IFuSol}), since values are positive, yields:
\begin{equation}\label{giFl}
 v(t) \leq v_{\lor}(t) \deq (v(t_0) - v_\circ) \, e^{-\frac{t - t_0}{\tau_{\land}}} + v_\bullet + i_\bullet \, t
\end{equation}
writing:
\begin{equation}\begin{array}{rcl} \label{giFl2}
i_\bullet &\deq& \tau_{\land}(t_0, t_1) \, i'_{\lor}(t_0 , t_1) \\
v_\bullet &\deq& \tau_{\land}(t_0, t_1) \, (i_{\lor}(t_0 , t_1) - \tau_{\land}(t_0, t_1) \, i'_{\lor}(t_0 , t_1)) \\
v_\circ   &\deq& v_\bullet + i_\bullet \, t_0 \\
\end{array} \end{equation}

Finally we can solve the equation for $t^{\lor} \deq v_{\lor}^{-1}(1)$ and obtain:
\begin{equation}\label{giFt}
t^{\lor}(t_0, t_1) = \left\{\begin{array}{lcl}
t_0 &,& v(t0) \geq 1 \\
\frac{1 - v_\bullet}{i_\bullet} + \tau_{\land} \, \mbox{L}\left( \frac{v_\circ - v(t_0)}{\tau_{\land} \, i_\bullet} \, e^{\frac{v_\circ - 1}{\tau_{\land} \, i_\bullet}}\right)
&,& i'_{\lor}(t_0 , t_1) > 0 \\
t_0 + \tau_{\land} \, \log\left(\frac{v_\bullet - v(t_0)}{v_\bullet - 1}\right)
&,& i'_{\lor}(t_0 , t_1) = 0, v_\bullet > 1 \\ 
+\infty &,& \mbox{ otherwise } \\ 
\end{array} \right. \end{equation}

Here $y = \mbox{L}(x)$ is the Lambert function defined as the solution, analytic at 0, of $y \, e^y = x$ and is easily tabulated.

The derivations details are omitted since they have been easily obtained using a symbolic calculator.

In the case where $i'_{\lor}(t_0 , t_1) = 0$ thus $i_\bullet = 0$, $v_{\lor}(t)$ corresponds to a simple leaky integrate and fire  neuron (LIF) dynamics 
and the method thus consists of upper bounding the gIF dynamics by a LIF in order to estimate a spiking time lower-bound. 
This occurs when constant upper-bounds is used for the currents. Otherwise~(\ref{giFl}) and~(\ref{giFl2}) corresponds to more general dynamics.

\subsection{Event-based iterative solution}

\begin{figure}[htbp]
\centerline{\includegraphics[width=0.45\textwidth]{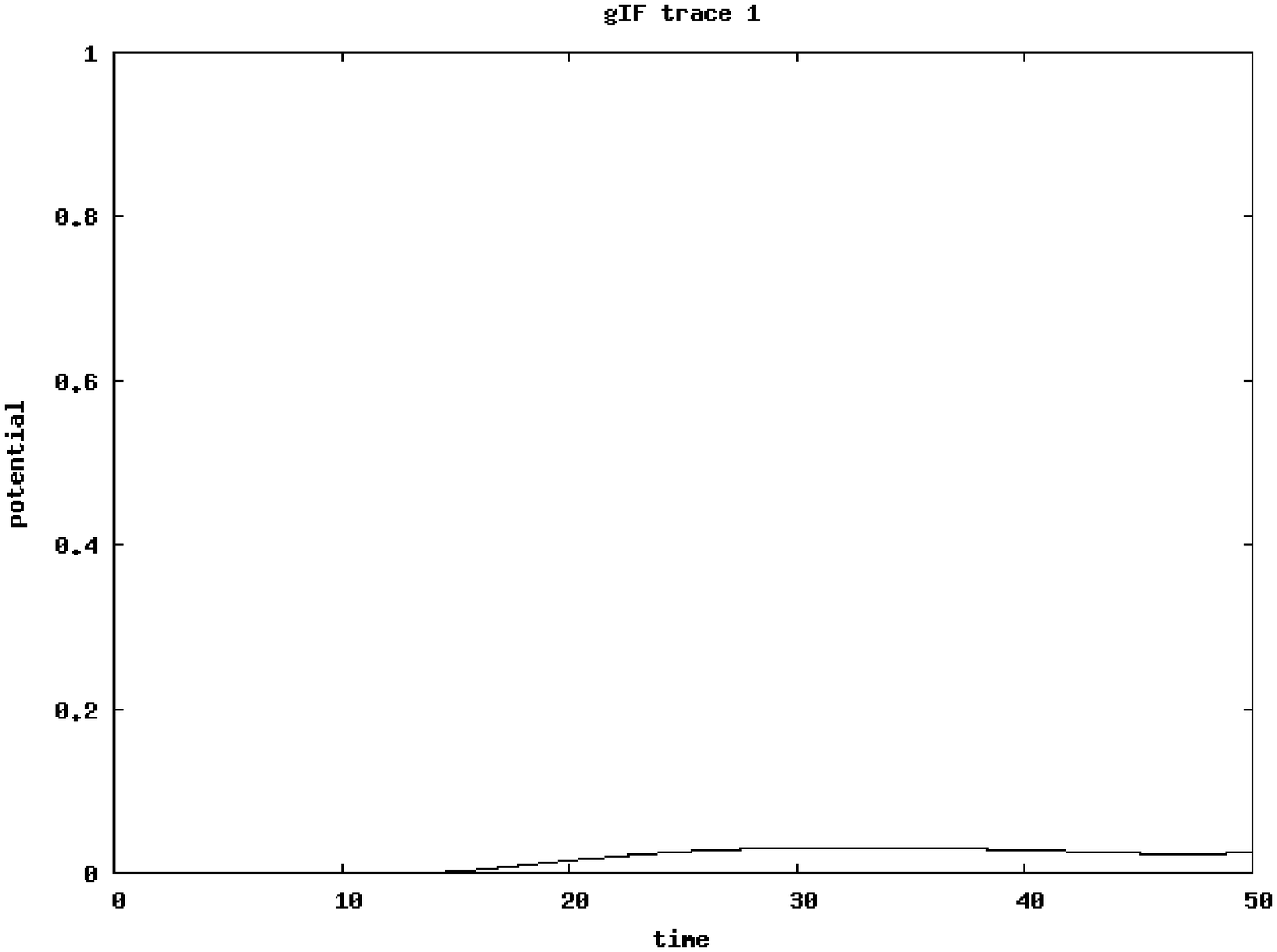}~~~~\includegraphics[width=0.45\textwidth]{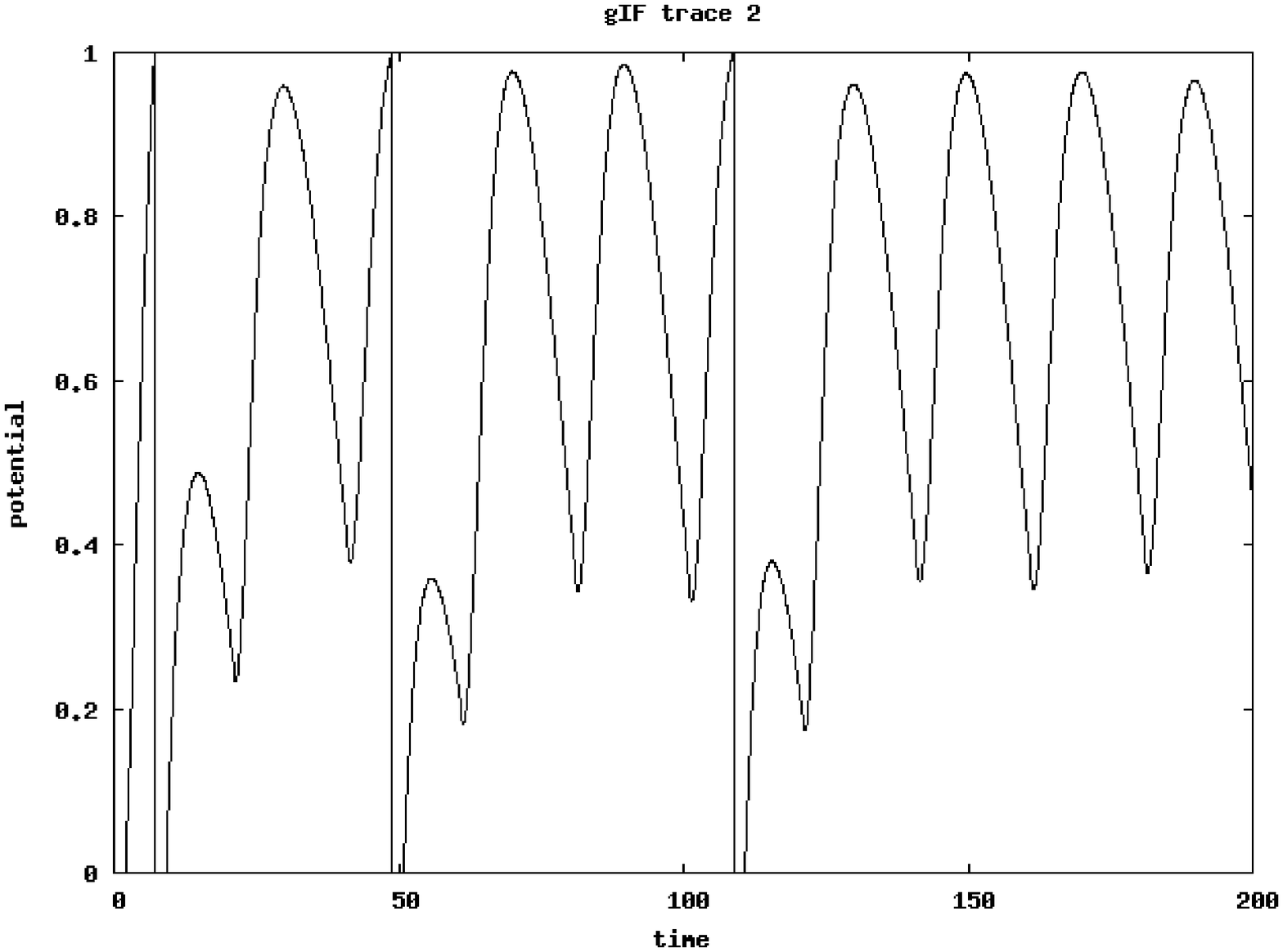}}
\caption{\label{gif-trace} An example of gIF normalized membrane potential.
The left trace corresponds to $50ms$ of simulation, 
the neuron being artificially connected to a periodic excitatory/inhibitory input neuron pair at 50/33 Hz of high synaptic weight,
in order to make explicit the double exponential profiles.
The right trace corresponds to $200ms$ of simulation, with a higher inhibition.
The weights have been chosen, in order to make the neuron adaptation explicit: the firing frequency decreases until obtaining a sub-threshold membrane potential.}
\end{figure}

Let us apply the previous derivation to the calculation of the next spike-time lower-bound for a gIF model, up to a precision $\epsilon_t$. 
One sample run is shown Fig.~\ref{gif-trace}

Given a set of spike-times $t_j^n$ and an interval $[t_0, t_1]$, from~(\ref{gIGlb1}) $\tau_{\land}$, $i_{\lor}$ and $i'_{\lor}$ are calculated in 
about $10^1\,M\,C$ operations, for an average of $C$ connections per units, and with an average number $M \leq N$ of firing units.
This is the costly part of the 
calculation\footnote{It appears, that since each synapse response corresponds to a linear differential equation which could be analytically integrated, 
the global synaptic response $\bar{r}_j(t) = \sum_n r_j(t - t_j^n)$ can be put in closed form, and then bounded by constant values,
thus reducing the computation complexity from $O(MC)$ to $O(C)$ as detailed in \cite{rotter-diesmann:99}. 
This well-known issue is not re-addressed here, simply to avoid making derivations too heavy.}, and is equivalent to a single clock-based integration step. 
Spike response profiles $r_j()$ and profile derivatives $r'_j()$ for excitatory/inhibitory synapses and gap junctions are tabulated with a $\epsilon_t$ step. Then,
from~(\ref{giFl2}) and~(\ref{giFt}) we obtain $t^\lor(t_0, t_1)$.

The potential $v(t_0)$ is calculated using any well-established method, as detailed in, e.g., \cite{rotter-diesmann:99}, and not reviewed here.

The following algorithm guarantees the estimation of a next spike-time lower-bound after $t_0$. Let us consider an initial interval of estimation $d$, say $d \simeq 10ms$:
\begin{quotation}
 -a- The lower-bound $t_\lor = t^\lor(t_0, t_0 + d)$ is calculated.

 -b- If $t_0 + d \leq t_\lor$ 
	\begin{tabular}{ll} the lower-bound value $t_0 + d$ is returned \\ the estimation interval is doubled $d \leftarrow 2 \, d$ \end{tabular}

 -c- If $t_0 + \epsilon_t < t_\lor < t_0 + d$ 
	\begin{tabular}{ll} the lower-bound value $t_\lor$ is returned \\ the estimation interval is reduced $d \leftarrow 1/\sqrt{2} \, d$ \end{tabular}

 -d- If $t_0  < t_\lor < t_0 + d$, $d \leq \epsilon_t$, the next spike-time $t_\lor$ is returned.
\end{quotation}

Step -b- corresponds to the case where the neuron is not firing in the estimation interval. 
Since  $v(t)$ is bounded by $v_\lor(t)$ and the latter is reaching the threshold only outside $[t_0, t_0 + d]$, $t_0 + d$ is a time lower-bound. 
In addition, a heuristic is introduced, in order to increase the estimation interval in order to save computation steps.

Step -c- corresponds to a strict lower-bound computation, with a relative value higher than the precision $\epsilon_t$.

Step -d- assumes that the lower-bound estimation converges towards the true spike-time estimation when $t_1 \rightarrow t_0$. 

This additional convergence property is easy to derive. Since:
\\ \centerline{$\lim_{t_1 \rightarrow t_0} r_{j\land}(t_0, t_1) =  \lim_{t_1 \rightarrow t_0} r_{j\lor}(t_0, t_1) =  r_{j}(t_0), 
                \lim_{t_1 \rightarrow t_0} r'_{j\lor}(t_0, t_1) =  r'_{j}(t_0)$,}\\
then:
\\ \centerline{$\lim_{t_1 \rightarrow t_0} 1 / \tau_{\land}(t_0, t_1) = g(t_0), 
                \lim_{t_1 \rightarrow t_0} i_{\lor}(t_0, t_1) =  i(t_0),
                \lim_{t_1 \rightarrow t_0} i'_{\lor}(t_0, t_1) =  i'(t_0)$,} \\
yielding:
\\ \centerline{$\lim_{t_1 \rightarrow t_0} v_{\lor}(t) = \bar{v}_{\lor}(t) \deq (v(t_0) - \bar{v}_\circ)\,e^{-g(t_0)\,(t - t_0)} +  \bar{v}_\bullet + \bar{i}_\bullet \, t$} \\
writing:
\\ \centerline{$\begin{array}{rcl}
\bar{i}_\bullet &\deq& 1/g(t_0) \, i'(t_0) \\
\bar{v}_\bullet &\deq& 1/g(t_0)\, (i(t_0) - 1/g(t_0) \, i'(t_0)) \\
\bar{v}_\circ   &\deq& \bar{v}_\bullet + \bar{i}_\bullet \, t_0 \\
\end{array}$}
We thus obtain a limit expression $\bar{v}_{\lor}$ of $v_{\lor}$ when $t_1 \rightarrow t_0$. From this limit expression we easily derive:
\\ \centerline{$v(t) - \bar{v}_{\lor}(t) = -1/2 \, g'(t_0) \, v(t_0) \, t^2 + O(t^3)$,}
and finally obtain a quadratic convergence, with an error closed-form estimation.

The methods thus corresponds to a semi-interval estimation methods of the next spike-time, the precision $\epsilon_t$ being adjustable at will. 

The interval of estimation $d$ is adjusted by a very simple heuristic, which is of standard use in non-linear numerical calculation adjustment.

The unit calculation corresponds to one step of the iterative estimation, the estimation loop being embedded in the simulator interactions.
This is an important property as far as real-time computation is concerned, 
since a minimal amount of calculation is produced to provide, as soon as possible, as suboptimal answer.

This ``lazy'' evaluation method is to be completed by other heuristics: 
\\ - if the input spike is inhibitory thus only delaying the next spike-time, re-calculation can be avoided;
\\ - if all excitatory contributions $g_\lor = M \, r^+(t_b)$ are below the spiking threshold, spike-time is obviously infinity;
\\ - after a spike the refractory period allows us to postpone all calculation (although synaptic conductances still integrate incoming spikes).

In any case, comparing to other gIF models event-based simulation methods \cite{brette:06,brette:07,rudolph-destexhe:06}, this alternative method allows one to control 
the spike-time precision, does not constraint the synaptic response profile and seems of rather low computational cost, due to the ``lazy'' evaluation mechanisms.

\subsubsection*{Numerical convergence of the lower-bound iteration}

Considering biologically plausible parameters as reviewed here and in appendix~\ref{gif-normalization}, we have experimented carefully the numerical convergence 
of this lower-bound iterative estimation, considering a gIF neuron with adaptive and non-linear internal currents, implemented as proposed here, and providing 
membrane potential, e.g., as shown in Fig.~\ref{gif-trace}. 

Let us report our numerical experimentation.

We have always observed the convergence of the method (also extensively experimented at the network level in a next section), with a convergence in about $2-20$ iterations 
(mean $\simeq 11$, standard-deviation $\simeq 5$), the lower-bound iteration generating steps of about $0.01-10ms$ (mean $\simeq 3ms$, standard-deviation $\simeq 4ms$) 
from one lower-bound to another, with three distinct qualitative behaviors:
 \\ -a- Sub-threshold maximal potential: the previous calculation estimates a maximal membrane potential below the threshold and calculation stops,
the neuron being quiet; in this mode the event-based strategy is optimal and a large number of calculations are avoided with respect to clock-based paradigms.
 \\ -b- Sub-threshold lower-bound estimation: the maximal membrane potential is still estimated over the threshold, with a next-spike time lower bound. In this mode,
we observed an exponential increase of the next-spike time lower bound and in $2-5$ iterations the maximal membrane potential estimation is estimated under the threshold,
switching to mode -a-; in this mode the estimation interval heuristic introduced previously is essential and the next-spike time lower bound estimation allows the calculation
to quickly detect if the neuron is quiet.
 \\ -c- Iterative next-spike time estimation: if a spike is pending, the previous calculation estimates in about $10-20$ iterations the next-spike time up to a tunable
precision (corresponding to the $d t$ of the simulation mechanism). The present mechanism acts as an iterative estimation of the next spike time, as expected.

\section{About event-based simulation of SRM models}\label{srm-simul}

\begin{figure}[htbp]
\centerline{
\begin{tabular}{c@{\hspace*{2cm}}c}
\includegraphics[width=0.29\columnwidth,height=1.5cm]{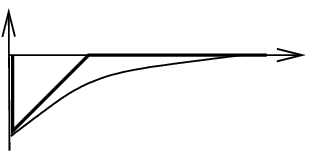}&
\includegraphics[width=0.29\columnwidth,height=1.5cm]{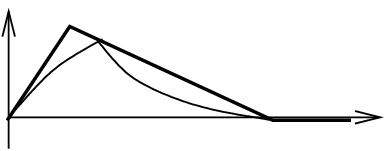}\\
$\nu_i$ & $\varepsilon_{ij}$
\end{tabular}
}
\caption{\label{fig:someFunctions} Potential profiles used in equation (\ref{eq:srm}). The original exponential profiles derived by the authors of the model are shown
as thin curves and the piece-wise linear approximation as thick lines. 
Any other piece-wise linear profiles could be considered, including piece-wise finer linear approximation of exponential profiles.}
\end{figure}

 Among spiking-neuron models, the Gerstner and Kistler spike response model (SRM) \cite{gerstner-kistler:02b} of a biological neuron defines the state of a neuron via a 
single variable: \begin{align}\label{eq:srm}
 u_i(t) = r_i \, j(t) + \nu_i(t - t_i^*) +    \sum_{j} \sum_{t_j^n \in {\cal F}_j} w_{ij} \, \varepsilon_{ij}(t - t_i^*, (t - t_j^n) - \delta_{ij}) 
\end{align}
where $u_i$ is the normalized membrane potential, 
$j()$ is the continuous input current for an input resistance $r_i$. 
The neuron fires when $u_i(t) \geq \theta_i$, for a given threshold,
$\nu_i$ describes the neuronal response to its own spike (neuronal refractoriness), 
$t_i^*$ is the last spiking time of the $i$th neuron, 
$\varepsilon_{ij}$ is the synaptic response to pre-synaptic spikes at time $t_j^n$ post-synaptic potential (see Fig.~\ref{fig:someFunctions}), 
$w_{ij}$ is the {\em connection strength} (excitatory if $w_{ij} > 0$ or inhibitory if $w_{ij} < 0$) and 
$\delta_{ij}$ is the {\em connection delay} (including axonal delay). 
Here we consider only the last spiking time $t_i^*$ for the sake of simplicity, 
whereas the present implementation is easily generalizable to the case where several are taken into account.

Let us call here lSRM such piece-wise linear approximations of SRM models.

This model is very useful both at the theoretical and simulation levels. 
At a computational level, it has been used (see \cite{maass-bishop:03} for a review) to show
that any feed-forward or recurrent (multi-layer) analog neuronal network (\`a-la Hopfield, e.g., McCulloch-Pitts) 
can be simulated arbitrarily closely by an insignificantly larger network of spiking neurons, even in the presence of noise,
while the reverse is not true \cite{maass:97,maass-natschlager:97}.
In this case, inputs and outputs are encoded by temporal delays of spikes. 
These results highly motivate the use of spiking neural networks.

This lSRM model has also been used elsewhere (see e.g. \cite{schrauwen:07} for a review), 
including high-level specifications of neural network processing related to variational approaches \cite{vieville-chemla-etal:07},
using spiking networks \cite{vieville-rochel:06}. The authors used again a lSRM to implement their non-linear computations. 

\begin{figure}[htbp]
\centerline{\includegraphics[width=0.45\textwidth]{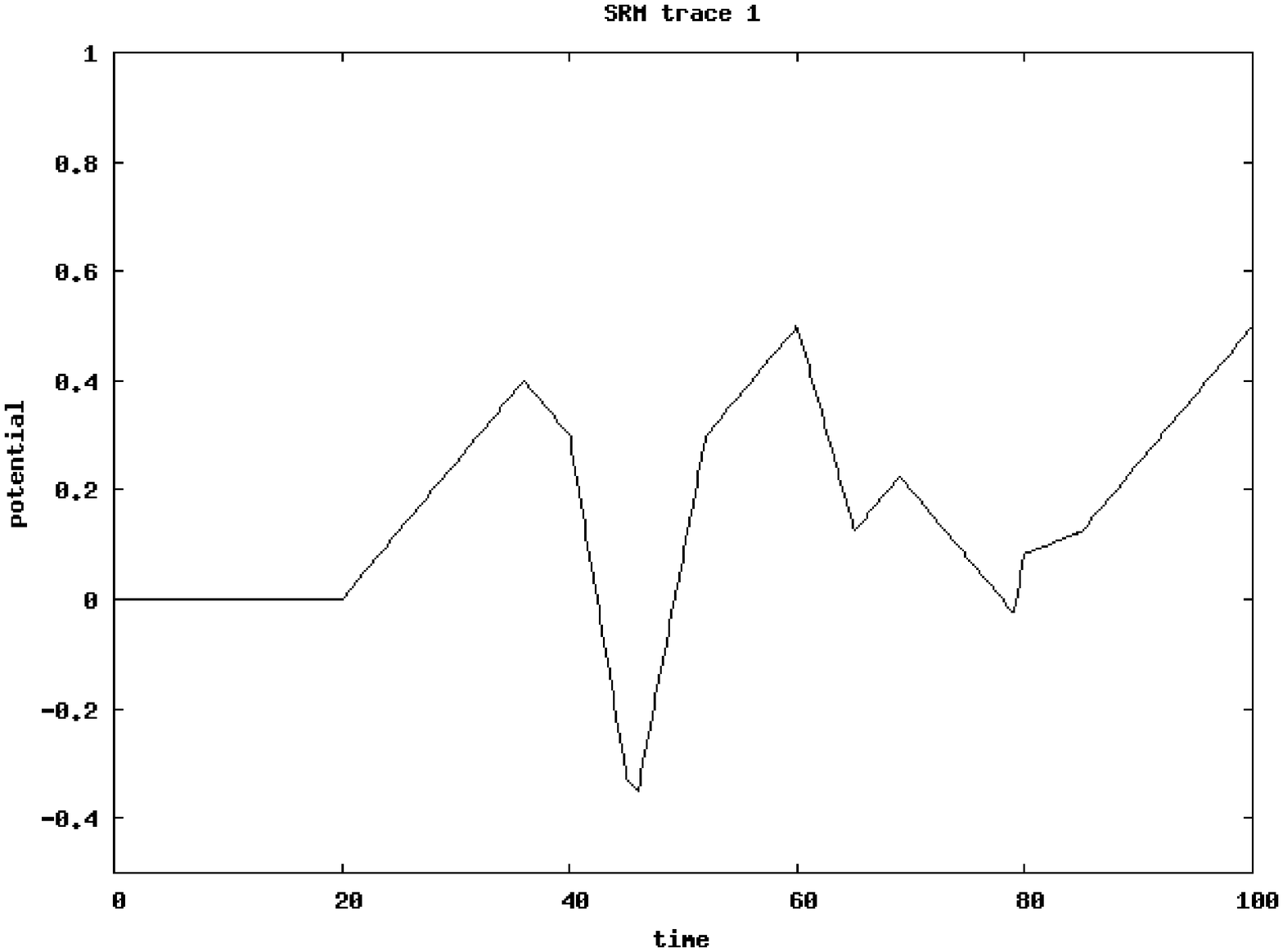}~~~~\includegraphics[width=0.45\textwidth]{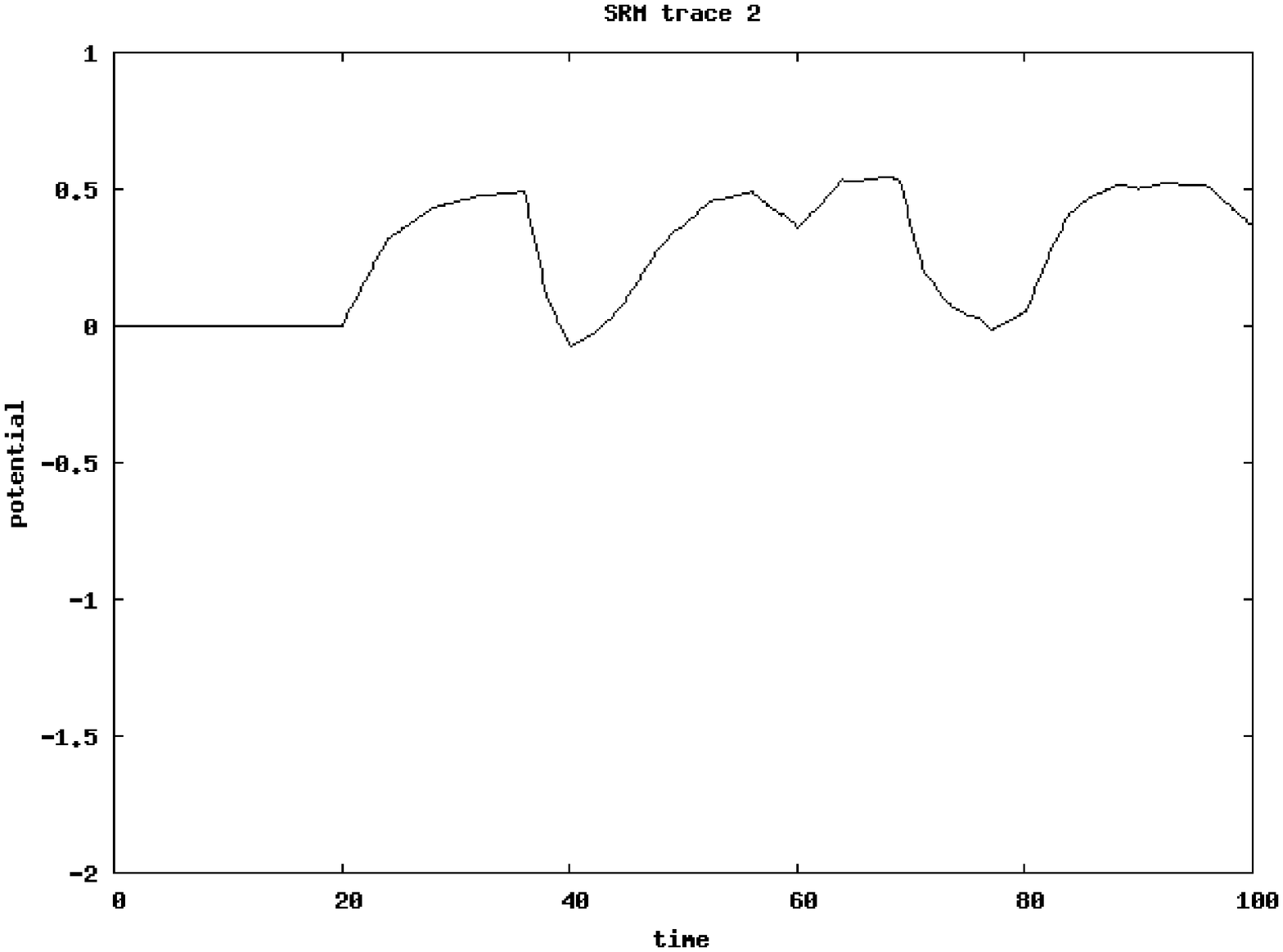}}
\caption{\label{srm-trace} An example of lSRM normalized membrane potential.
The traces corresponds to $100ms$ of simulation.
The leftward trace uses the fastest possible piece-wise linear approximation of the SRM model profiles.
The rightward trace is simulated with a lower excitation inhibition and using a thinner piece-wise linear approximation. 
The neuron is defined by biologically plausible parameters as reviewed in appendix~\ref{gif-normalization} .
It is connected to a pair of periodic excitatory and inhibitory input neurons, with different time constants, as in Fig.~\ref{gif-trace}.}
\end{figure}

Let us make explicit here the fact a lSRM can be simulated on an event-based simulator for two simple reasons:
\\ - the membrane potential is a piece-wise linear function as the sum of piece-wise linear functions 
  (as soon as the optional input current is also piece-wise constant or linear),
\\ - the next spike-time calculation is obvious to calculate on a piece-wise linear potential profile, 
  scanning the linear segments and detecting the 1st intersection with $u = 1$ if any.
The related piece-wise linear curve data-structure has been implemented$^{\mbox{\scriptsize \ref{enas}}}$ and support three main operations:
\\ - At each spike occurrence, add linear pieces to the curve corresponding to refractoriness or synaptic response.
\\ - Reset the curve, after a spike occurrence.
\\ - Solve the next spike-time calculation.
\\ This is illustrated in Fig.~\ref{srm-trace}.

This is to be compared with other simulations (e.g. \cite{mattia-giudice:00,ruf:98}) where stronger simplifications of the SRM models have been introduced to obtain 
a similar efficiency, whereas other authors propose heavy numerical resolutions at each step. 
When switching from piece-wise linear profiles to the original exponential profiles \cite{gerstner-kistler:02b}, the equation to solve is now of the formal form:
\\ \centerline{$1 = \sum_{i = 1}^n \lambda_i  e^{t/\tau_i}$,} \\ 
without any closed-form solution as soon as $n > 1$. 
One elegant solution \cite{brette:07} is to approximate the time-constant by rational numbers and reduce this problem to a polynomial root finding problem.
Another solution is to upper-bound the exponential profiles by piece-wise linear profiles in order to obtain a lower-bound estimation of the next spike-time and 
refine in the same way as what has been proposed in the previous section. Since the mechanism is identical, we are not going to further develop.

In any case, this very powerful phenomenological model of biological dynamics can be simulated with several event-based methods, including at a fine degree of precision,
using more complex piece-linear profiles.

\section{Experimental results} \label{results}
\subsection{Kernel performances and features} \label{performances}

In order to estimate the kernel sampling capability we have used, as a first test, a random spiking network with parameter less connections,
the spiking being purely random thus not dependent on any input.

\begin{figure}[htb]
 \centerline{\includegraphics[width=0.8\textwidth]{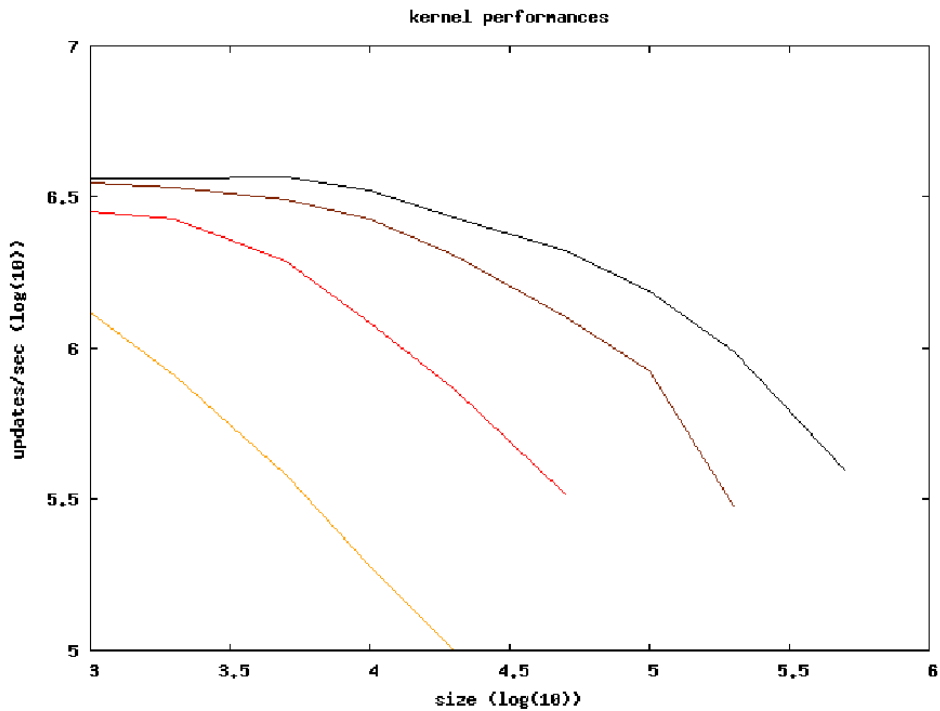}}
\caption{\label{perf} Simulation performances, for $10^5$ spike firings and about $0.1s$ of simulation time. 
 The CPU time is about $2s$ for $2^{13}$ neurons and $2^{21}$ synapses and does not depend on $D$ or $dt$ values, as expected.
 Spike-time insertion/deletion are counted as elementary updates.
 The network size in abscissa varies from $10^3$ to about $10^6$ and the number of connections from from 0 to $10^8$, corresponding to curve end-points.  
 Curves are shown for a connection probability of $P = 0$ (black, upper curve), $P = 10^{-3}$ (brown), $P = 10^{-2}$ (red), $P = 10^{-1}$ (orange). 
 The performance is mainly function of the number of synapses, and marginally of the number of neurons.}
\end{figure}

In term of performances, on a standard portable computer ({\small Pentium M 750 1.86 GHz, 512Mo of memory}) we process about $10^{5-7}$
spike-time updates / second, given the network size and connectivity. 
Performances reported in Fig.~\ref{perf} confirm that the algorithmic complexity only marginally depends on the network size, while it is
mainly function of the number of synapses (although both quantities are indeed linked).
We also notice the expected tiny overhead when iterating on empty boxes in the histogram, mainly visible when the number of spikes is small. 
This overhead is constant for a given simulation time.
The lack of proportionality in performances is due to the introduction of some optimization in the evaluation of spike-times, which are not updated if unchanged.

We have also observed that the spike-time structure upper and lower bounds $D$ and $dt$ have only a marginal influence on the performances, as expected.

Moreover, the numbers in Fig.~\ref{perf} allows to derive an important number: the overhead for an event-based implementation of a clock-based mechanism. 
Since we can process about $2\,10^{6}$ updates/second while we have measured independently that a minimal clock-based mechanism process about $5\,10^{7}$ updates/second,
we see that the cost the overhead is of about $0.5\mu s$ / update. This number is in coherence with the number of operations to realize at time modification in
the underlying data structure.

It is important to clarify these apparently ``huge'' performances. The reason is that the event-based simulation kernel is {\em minimal}.
As detailed in Table~\ref{unit}, the implementation make a simple but extensive use of the best mechanisms of object oriented implementations. 
The network mechanism (i.e., the kernel) corresponds to about 10Kb of C++ source code, using a ${\cal O}(D/dt+N)$ buffer size and 
about ${\cal O}(1+C+\epsilon/dt) \simeq 10-50$ operations/spike, for a size $N$ network with $C$ connections in average, while $\epsilon \ll 1$.
This $\epsilon$ corresponds to the overhead when iterating on empty boxes in the histogram. 
We can use such a simple spike-time data structure because of the temporal constraints taken into account in our specifications.

As a consequence, not all spiking mechanisms are going to be simulated with this kernel: 
units with event-time intervals or input/output event delay below $dt$ are going to generate a fatal error;
units with inter-event intervals higher than $D$ are also going to defeat this mechanism 
(unless an extension of the present mechanism, discussed previously, is not implemented).
Note that despite these limitations the event-time accuracy itself is the not $dt$ but the floating point machine precision.

\begin{table}[htb]
 \centerline{\parbox{0.9\textwidth}{\scriptsize \tt template <class C> class Unit \{
\\  // Gets the next alter time: event time or a lower-bound of the next event time.
\\  inline virtual double getNext(double present-time);
\\
\\  // Called to update the unit state, when the next spiking time or a lower-bound occurs. 
\\  // Returns true if an event occurs, false it was a lower-bound
\\  inline virtual bool next(double present-time);
\\
\\  // Called when an input unit fires an event.
\\  // Returns true if the next alert time is to be updated, false otherwise.
\\  inline virtual bool add(int neuron-index, C\& connection-parameter, double present-time);
\\\};}}
\caption{\label{unit} Specification of an event-based unit (pseudo-code). 
 Each unit (neuron or group of neurons) specifies is next ``alert'' time and informs the network about event-occurrence.
 Lazy evaluation is implemented, at this level, via the fact that alert time is optional updated when receiving an event.
 The connection is templated in order for the kernel to be optimally recompiled for each kind of connection, while unit's mechanisms are inlined,
 allowing the compiler to eliminate code interface.
 The connection parameters is passed by reference, in order adaptation mechanisms to be implemented.
 See text for further details.}
\end{table}

\paragraph{Clock-based sampling in event-based environment.}

A step further, we have implemented a discretized version of a gIF network, called BMS, as detailed in \cite{cessac:08,cessac-vieville:08} (equations not reported here). 
The interest of this test is the fact that we can compare spike by spike an event-based and clock-based simulation since the latter is well-defined, 
thus without any approximation with respect to the former (see \cite{cessac-vieville:08} for details). 

\begin{figure}[htb]
 \centerline{\includegraphics[width=14cm]{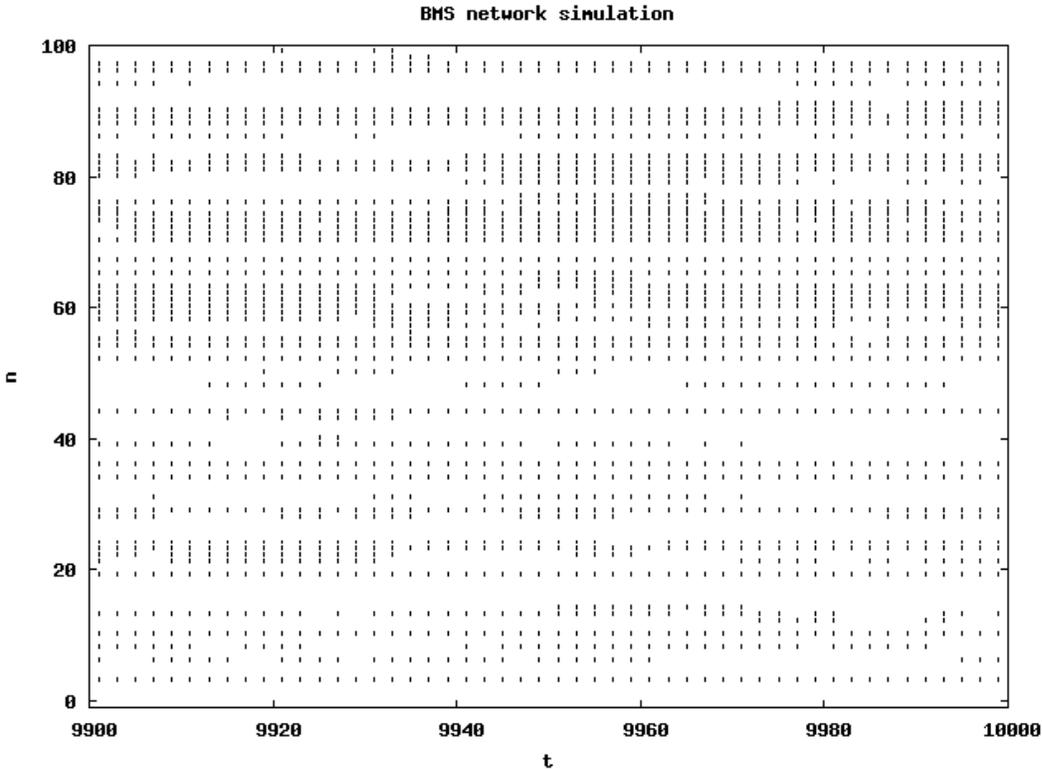}}
\caption{\label{BMS-network-simulation} An example of BMS neural network simulation used to evaluate the clock-based sampling in the event-based kernel, 
for $N = 1000$ and $T = 10000$ thus $10^{8}$ events. The first 100 neurons activity is shown at the end of the simulation.
The figure simply shows the strong network activity with the chosen parameters, and .}
\end{figure}

We have run simulation with fully connected networks of size, e.g. $N = 100-1000$ over observation periods of $T= 1000-10000$ clocks, 
with the same random initial conditions and the same randomly chosen weights, as show in Fig.\ref{BMS-network-simulation}. We have observed:

  -1- the same raster (i.e., with a Victor-Purpura distance of $0$ \cite{victor:05}); 
this exactitude is not surprising despite the fact that floating point errors accumulate\footnote{
Note that even if time is discretized, for BMS networks, the dynamics is based on floating point calculations, thus floating point errors accumulate.
However as soon as spike is fired, the potential is reset and previous errors are canceled. This explains why time-discretized simulations of IF networks
are numerically rather stable.}:
we are performing the same floating point calculations in both cases, 
since the event-based implementation is exact, thus . . with the same errors;

  -2- the overhead of the event-based implementation of the clock-based sampling is negligible (we obtain a number $< 0.1 \mu s/step$),
as expected. Again this surprisingly slow number is simply due to the minimal implementation, based on global time constraints, 
and the extensive use of the {\tt C/C++} optimization mechanisms.

\paragraph{Kernel usage.}

A large set of research groups in the field have identified what are the required features for such simulation tools \cite{brette-rudolph-etal:07}.
Although the present implementation is {\em not} a simulator, but a simple simulator plug-in, we have made the exercise to list to which extends what is proposed 
here fits with existing requirements, as detailed in Table~\ref{features}. We emphasize the fact the required programming is very light, 
for instance a ``clock'' neuron (allowing to mix clock-based and event-based mechanisms) writes:
\\\centerline{\parbox{0.9\textwidth}{\scriptsize \tt class ClockUnit : public Unit<bool> \{
\\  ClockUnit(double DT) : DT(DT), t(DT) { } double DT, t;
\\  inline virtual double getNext(double present-time) { return t; };
\\  inline virtual bool next(double present-time) { t += DT; return true; };
\\  inline virtual bool add(int neuron-index, bool connection-parameter, double present-time) { return false; };
\\\};}} \\
providing {\tt DT} is the sampling period. 
It is not easy to make things simple, and several possible choices of implementations have been investigated before proposing the interface proposed in Table~\ref{unit}.

See \cite{rochel:04} for a further description of how event-based spiking neuron mechanisms can be implemented within such framework. 
Although presented here at a very pragmatical level, 
note that these mechanisms are based on the modular or hierarchical modeling strategy borrowed from the DEVS formalism (see, e.g., \cite{rochel:04}).

\subsection{Experimenting reduced adaptive and ionic currents}

\begin{figure}[htb] 
 \centerline{\begin{tabular}{ccc}
   \includegraphics[width=4cm,height=2cm]{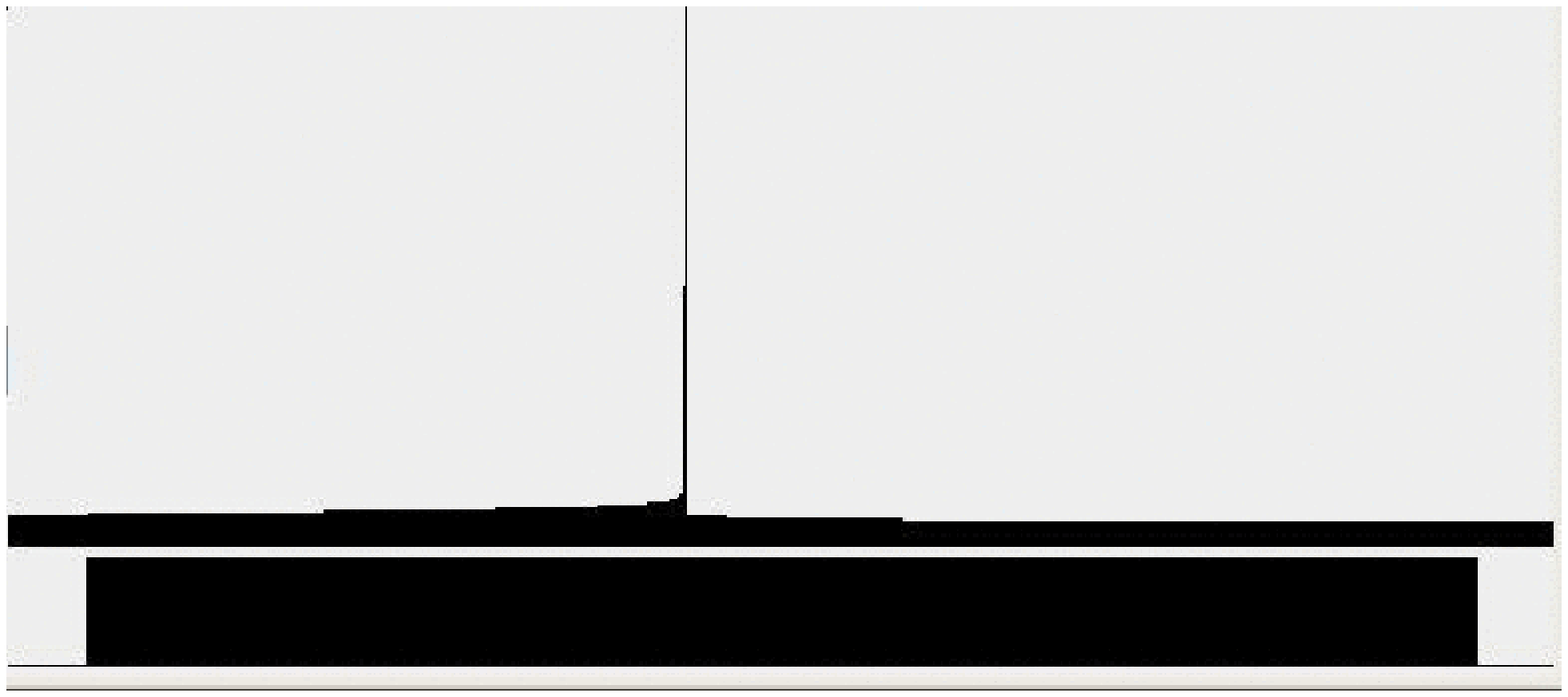} & \includegraphics[width=4cm,height=2cm]{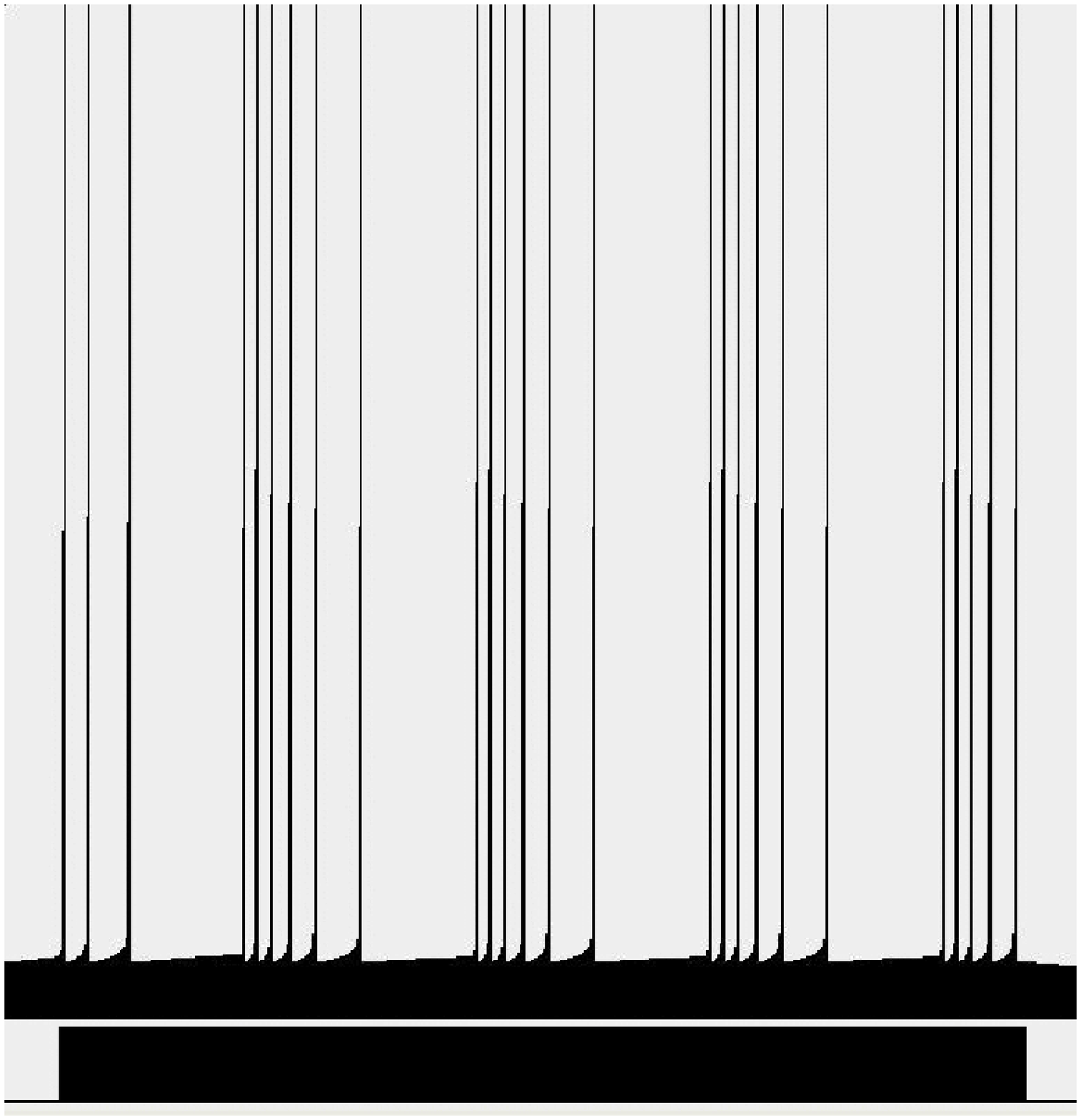} & \includegraphics[width=4cm,height=2cm]{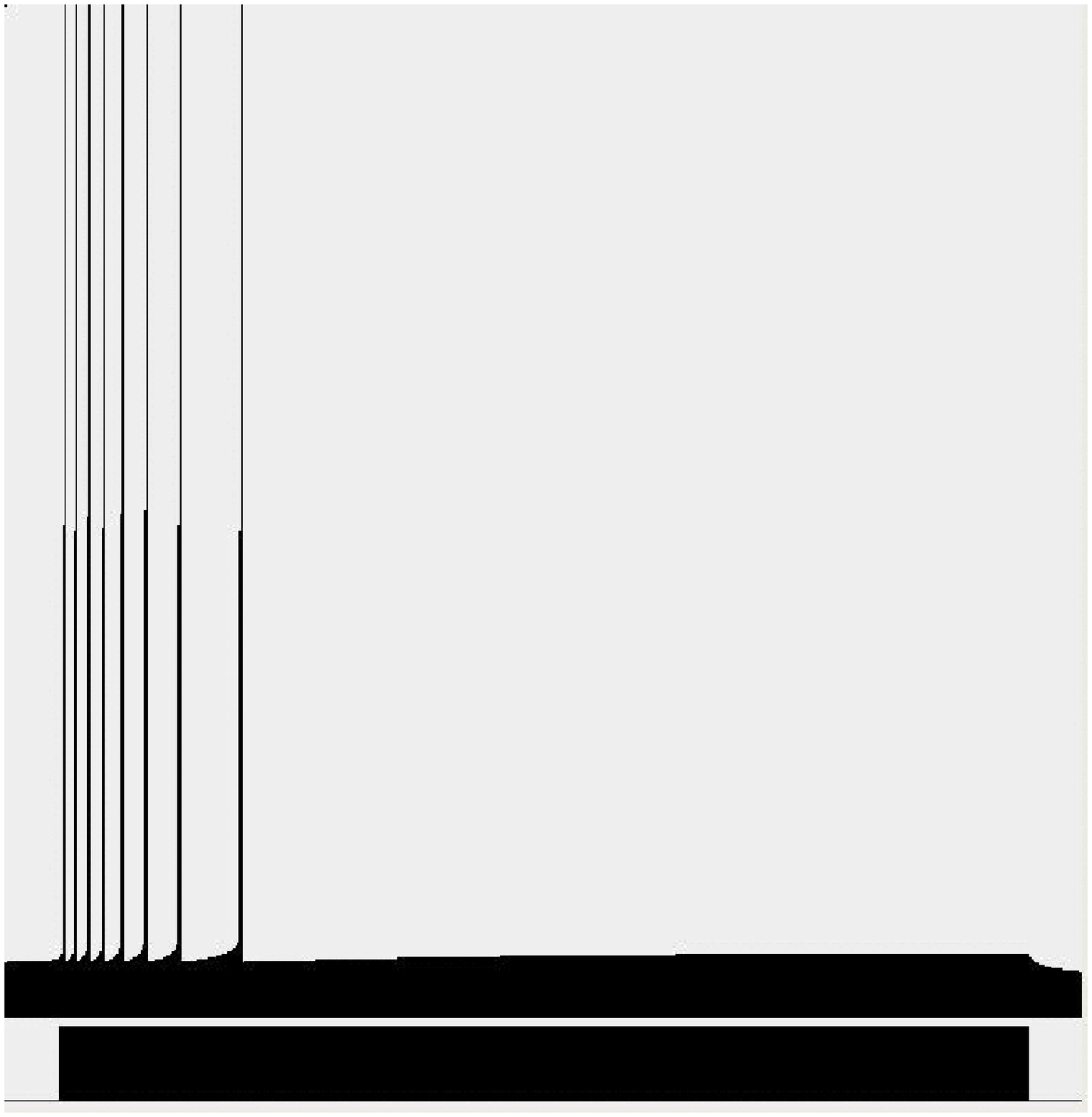} \\
    phasic-spiking & tonic-bursting & phasic-bursting \\
   \includegraphics[width=4cm,height=2cm]{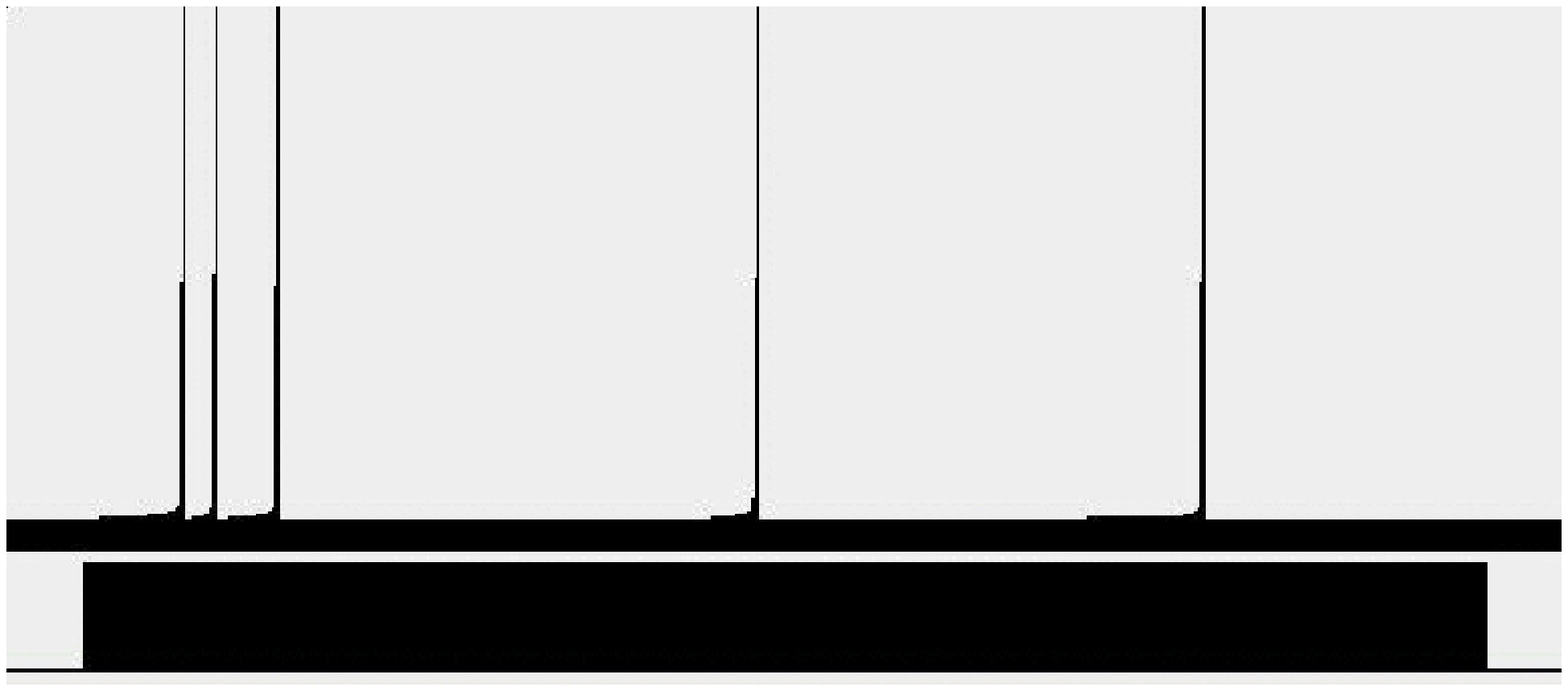} & \includegraphics[width=4cm,height=2cm]{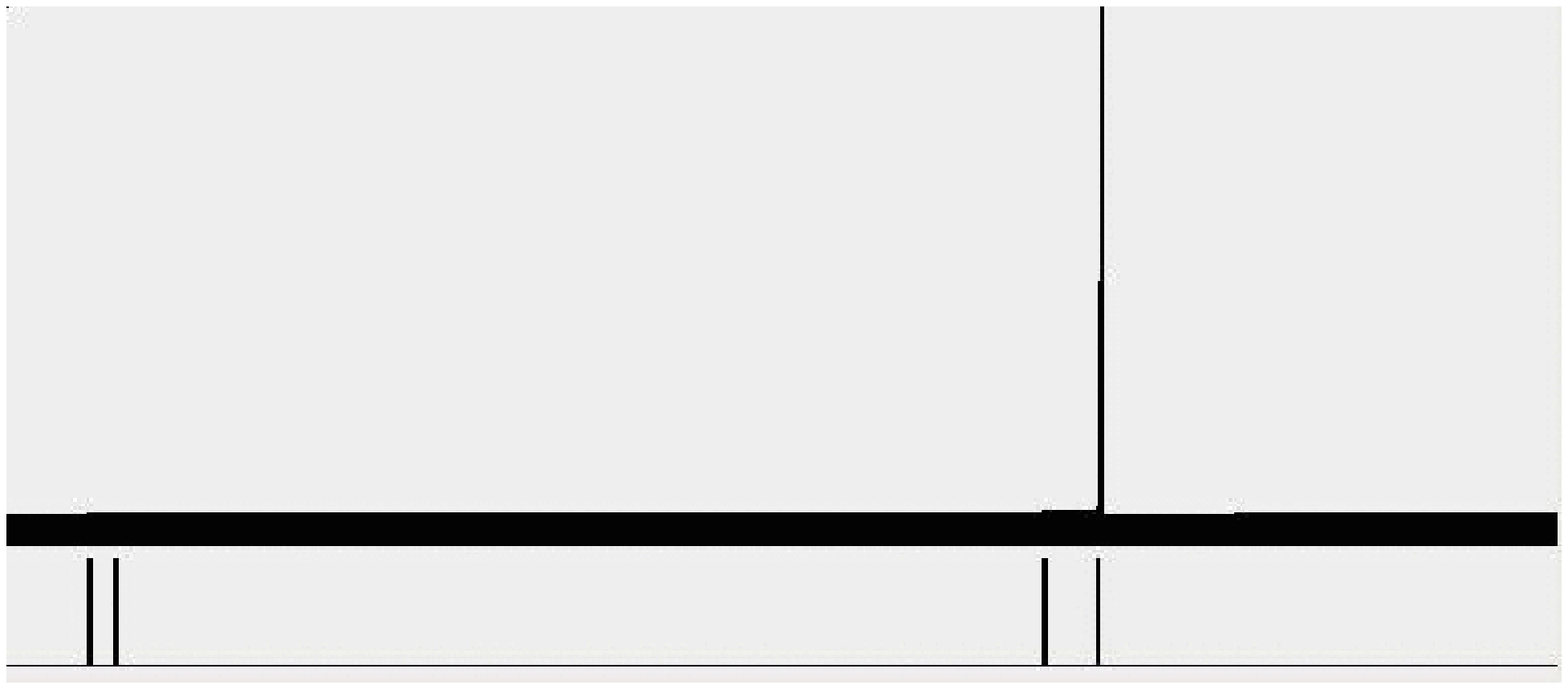} & \includegraphics[width=4cm,height=2cm]{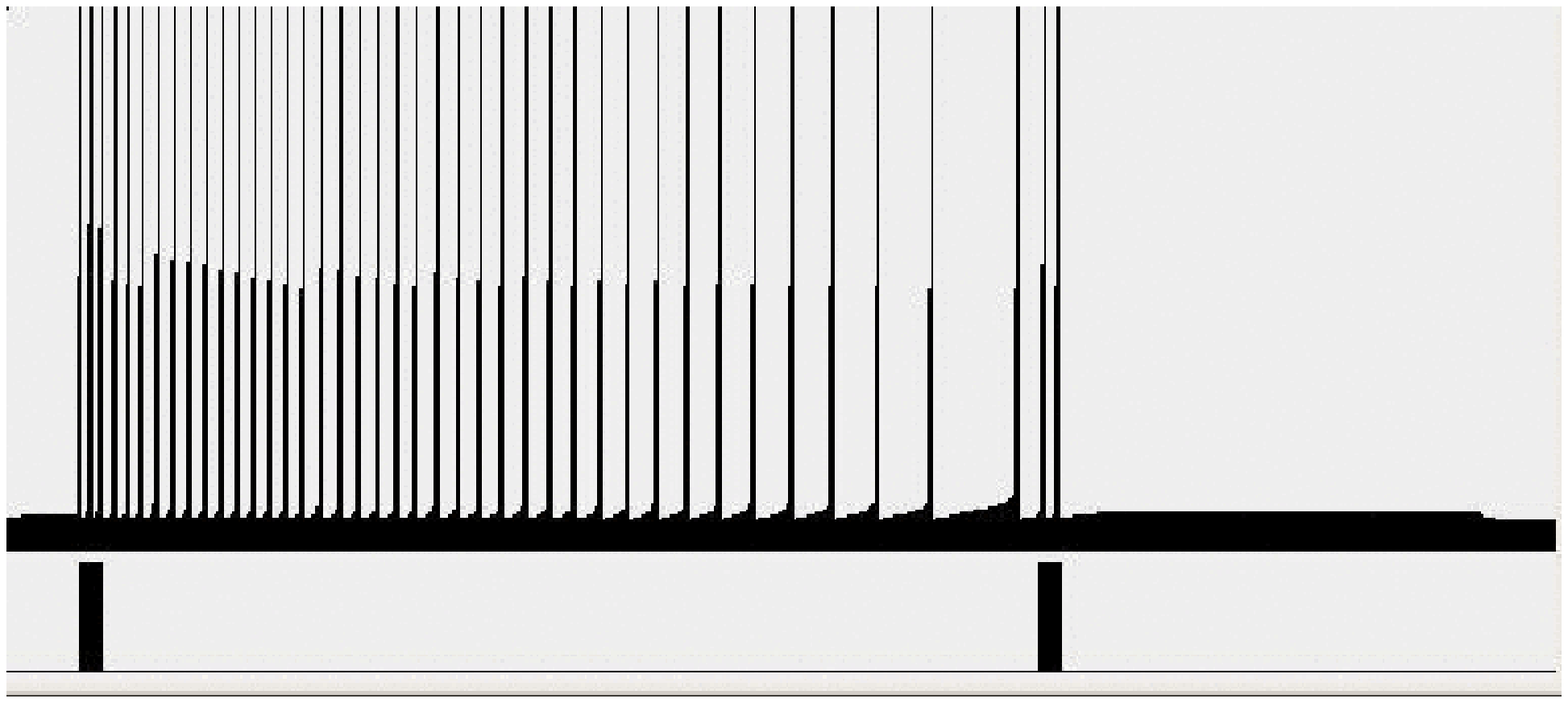} \\
    mixed-mode & resonator & bistability \\
   \includegraphics[width=4cm,height=2cm]{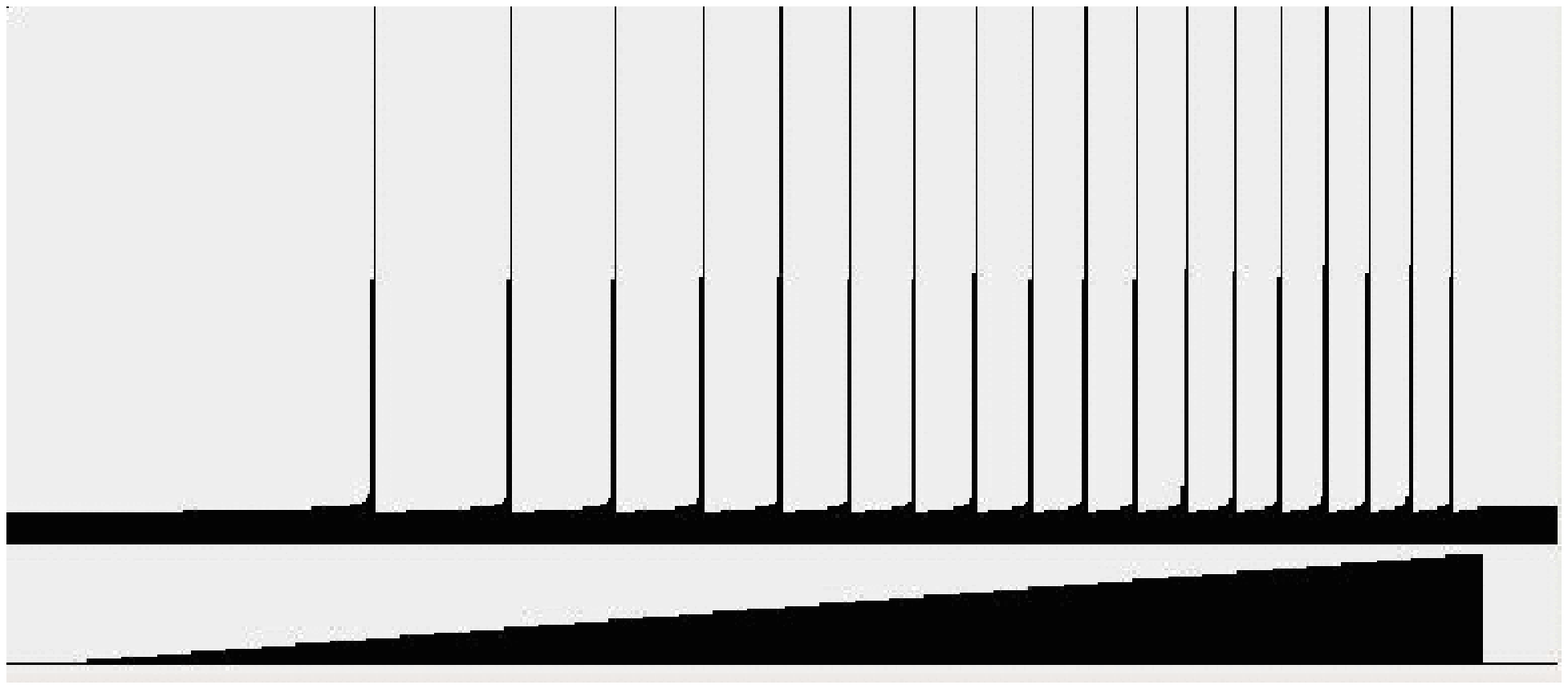} & \includegraphics[width=4cm,height=2cm]{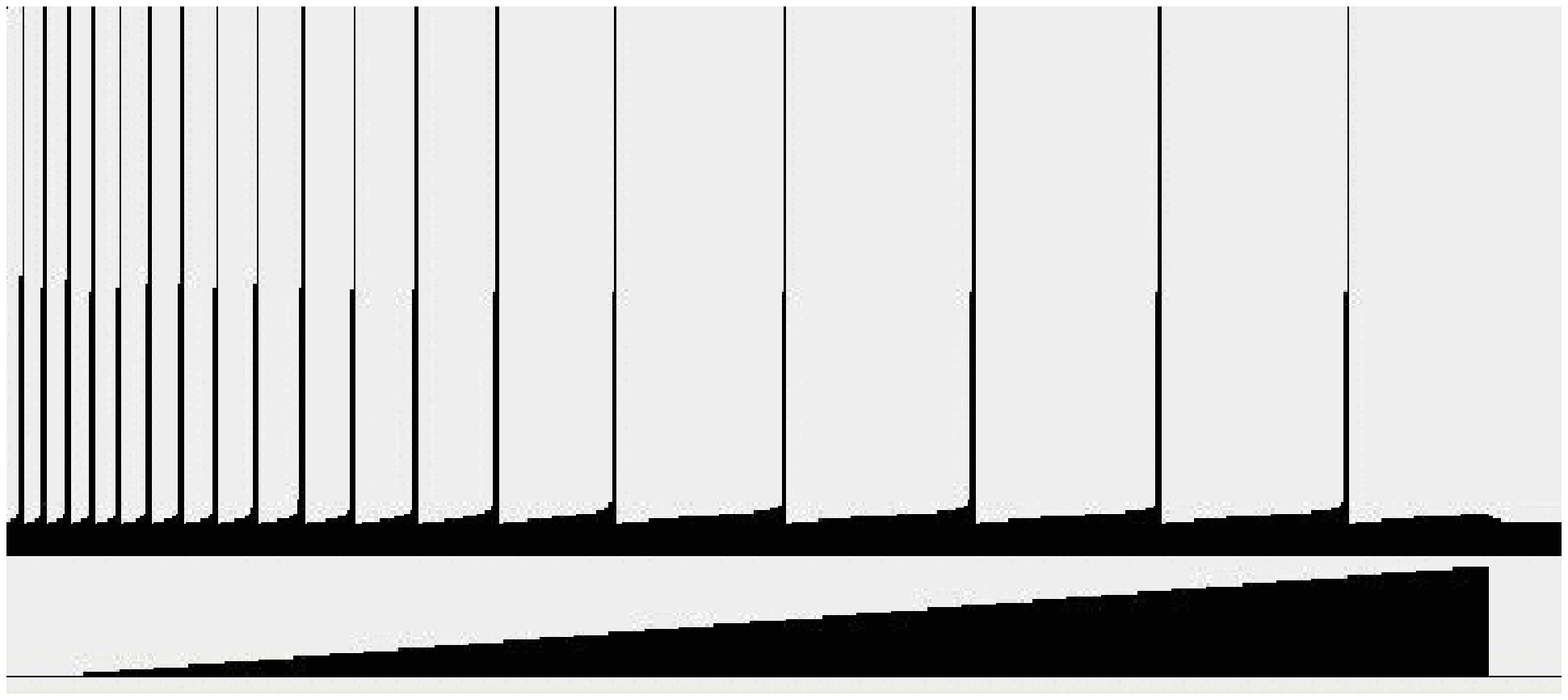} & \includegraphics[width=4cm,height=2cm]{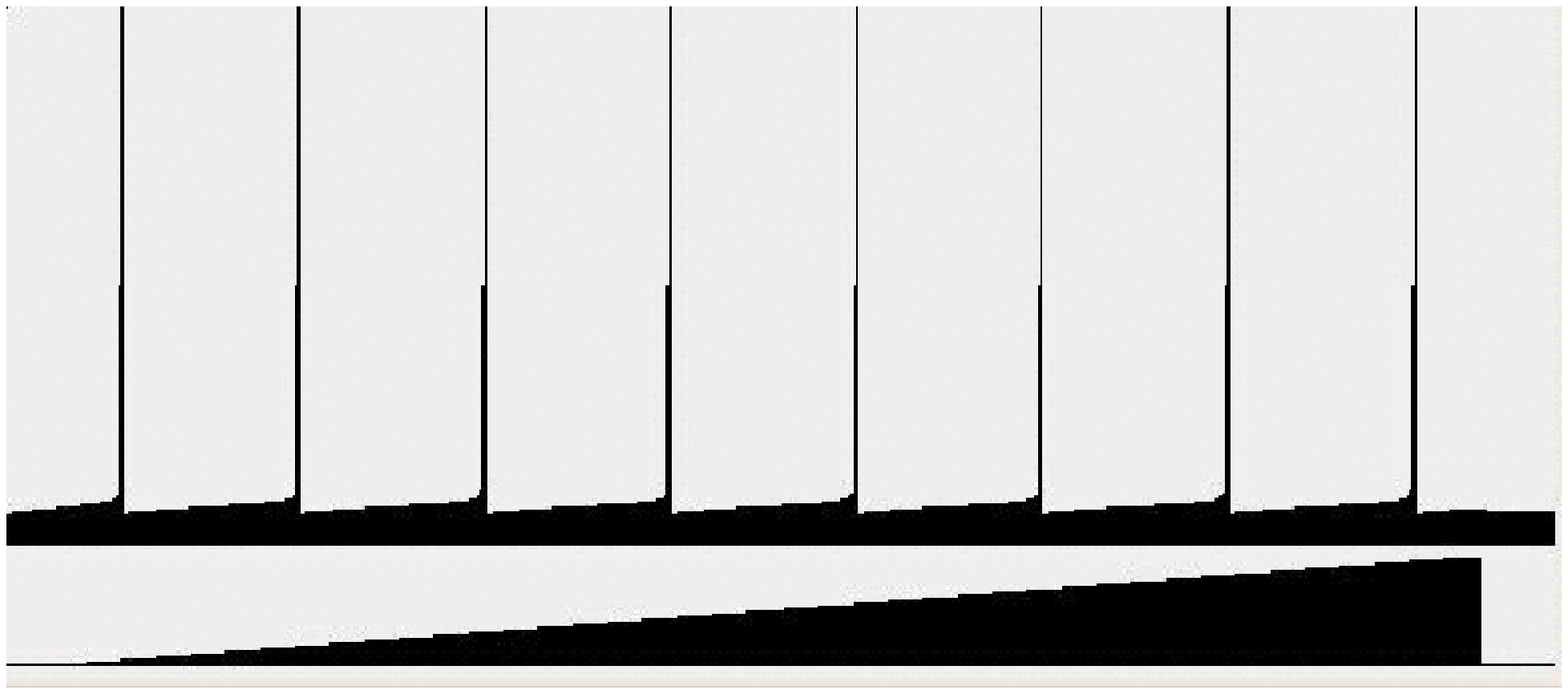} \\
     positive CFR & negative CFR & constant CFR \\
  \end{tabular}}
  \caption{\label{yvette-neurons} Typical results showing the versatility of the reduced model for spiking, bursting and other 
	modes, including and different current-frequency-responses (CFR). For each mode, the upper trace shows the action potentials, the lower trace the input current.
	These results include the excitatory mode of type I where the spike frequency can be as small as possible in a $1-10^3 Hz$ range and of type II where
        the spike frequency remains bounded. Tonic spiking is not shown, since obvious to obtain.}
 \end{figure}

 In order to experiment about our proposal to reduce ionic and adaptive currents to a function depending only on spike time, 
 we consider a very simple model whose evolution equation at time {\tt t} for the membrane potential {\tt v} is: 
 \begin{equation} \label{neuron-code} \mbox{\scriptsize
 \begin{tabular}{ll}
   {\tt if (t = 0)} \\
       & {\tt v = 0; u = 0; t\_0 = 0; } \\
   {\tt else if (v $\geq$ 1)} \\
       & {\tt v = 0; u = u + {\em k}; t\_0 = t; } \\
   {\tt else}  \\
       & {\tt $\dot{\tt v}$ = -{\em g} (v - {\em E}) - u +  {\em i}; } \\
       & {\tt if (t $>$ t\_0 + {\em d}) v = 0} \\
 \end{tabular} } \end{equation}
 where {\tt u} is the adaptive current (entirely defined by equation~(\ref{neuron-code})), {\tt t\_0} the last spiking time, $d$ the non-linear current delay. 
 The differential equation is simulated using an Euler interpolation as in \cite{izhikevich:03,touboul:08} to compare our result to what has been obtained by the other authors. 
 The obvious event-based simulation of this model has been also implemented$^{\mbox{\scriptsize \ref{enas}}}$.
 The input current {\tt \em i} is either a step or a ramp as detailed in Fig.~\ref{yvette-neurons} and Table~\ref{yvette-parameters}.

 Four parameters, the constant leak conductance {\tt \em g}, the reversal potential {\tt \em E}, the adaptation current {\tt \em k} step and the (eventually infinite) 
 non-linear current delay {\tt \em d} allows to fix the firing regime. These parameters are to be recalculated after the occurrence of each internal or external spike.
 In the present context, it was sufficient to use constant value except for one regime, as made explicit in Fig.~\ref{yvette-parameters}.
 We use the two-stages current whose action is to reset the membrane current after a certain delay. 
 We made this choice because it was the simplest and leads to a very fast implementation. 

 Experimental results are given in Fig.~\ref{yvette-neurons} for the parameters listed in Fig.~\ref{yvette-parameters}. 
 These results correspond to almost all well-defined regimes proposed in \cite{izhikevich:03}. 
 The parameter adjustment is very easy, we in fact use parameters given in \cite{izhikevich:03} with some tiny adjustments.
 It is an interesting numerical result: The different regimes are generated by parameters values closed to those chosen for the quadratic model, 
 the dynamic phase diagram being likely similar. See \cite{touboul:08} for a theoretic discussion.

 This places a new point on the performance/efficiency plane proposed by \cite{izhikevich:04} at a very challenging place, 
 and see that we can easily simulate different neuronal regimes with event-based simulations.

 However, it is clear that such model does {\em not} simulates properly the neuron membrane potential as it is the case for the exponential model \cite{brette-gerstner:05}.
 It is usable if and only if spike emission is considered, whereas the membrane potential value is ignored.

 \begin{table}[htb] 
   \centerline{\scriptsize \begin{tabular}{|l|c|c|c|c|c|c|}
   \hline
    spiking & leak        & reverse     & adaptation  & non-linear  & input    & input \\
    mode    & conductance & potential   & step        & delay       & magnitude  & form \\
            & {\tt \em g} & {\tt \em E} & {\tt \em k} & {\tt \em d} & \multicolumn{2}{c|}{\tt {\em i}(t)} \\
   \hline
    phasic-spiking   & 0.04 &   0 &  30  & $+\infty$ & 0.5 & step     \\
    tonic-bursting   & 0.18 & 1.6 & 14.6 &        60 &  15 & step     \\
    phasic-bursting  & 0.06 &  11 & 11.2 & $+\infty$ & 0.5 & step     \\
    mixed-mode       & 0.01 &   0 &  $K$ &       150 &  10 & step     \\
    resonator        & 0.04 & -27 &    0 & $+\infty$ &  38 & bi-pulse \\
    bistability      & 0.88 &  80 &  1.8 & $+\infty$ &  65 & pulse    \\
    positive CFR     & 0.01 &   0 &    0 & $+\infty$ &  30 & ramp     \\
    negative CFR     & 0.52 &  80 &    4 & $+\infty$ &  30 & ramp     \\
    constant CFR     & 0.52 &   0 &    4 &       100 &  30 & ramp     \\
  \hline \end{tabular}}
   \caption{\label{yvette-parameters} Examples of parameters used to generate the spiking modes shown in Fig.~\ref{yvette-neurons}.
   The mixed mode is simulated by a variable adaptation step $k = \{-20, 20\}$.}
 \end{table}

\subsection{Experimental benchmarks}\label{benchmarks}

\begin{figure}[htb]
 \centerline{\includegraphics[width=14cm,height=4cm]{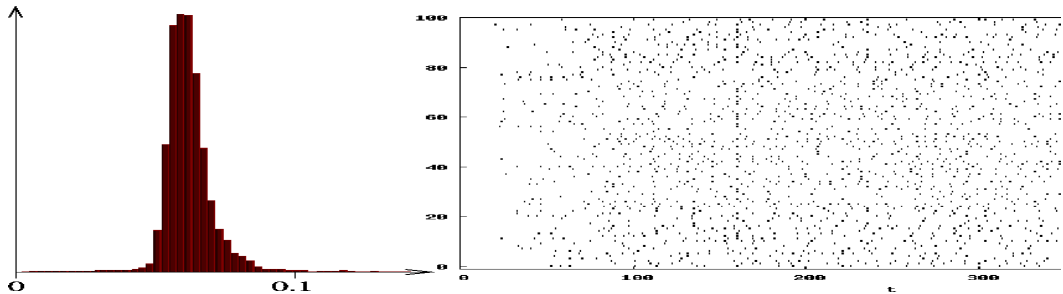}}
\caption{\label{gif-show} Inter-spike interval histogram (left view) in linear coordinates measured after $t > 0.1s$,
 and corresponding raster plot (right-view) during $1 s$ of simulation, for the benchmark 4 proposed in Appendix 2 of (Brette etal, 2007).}
\end{figure}

We have reproduced the benchmark 4 proposed in Appendix 2 of \cite{brette-rudolph-etal:07} which is dedicated to event-based simulation:
it consists of $4000$ IF neurons, which $80 / 20\%$ of excitatory/inhibitory neurons, connected randomly using a connection probability of $1/32 \simeq 3\%$. 
So called ``voltage-jump'' synaptic interactions are used: the membrane potential is abruptly increased/decreased by a value of 
$0.25 / 2.25 mV$ for each excitatory/inhibitory event (thus using fixed randomly chosen weights). 
Here, we also introduce a synaptic delay of $2 / 4 ms$ respectively and an absolute refractory period
of $1 ms$, both delays being corrupted by an additive random noise of $10\mu s$ of magnitude. We also have increased the network size and 
decreased the connection probability to study the related performances. In this network a synapse is simply defined by an index, weights are constant.
See \cite{brette-rudolph-etal:07} for further details. One result is proposed in Fig.~\ref{gif-show} to qualitatively verify the related network dynamics.
The fact we find small inter-spike intervals in this case is coherent with previous observed results.

A step further, we also made profit of the new proposed approximation of gIF neuron models to run another test, inspired by another benchmark 
proposed in \cite{brette-rudolph-etal:07}, after \cite{vogels-abbott:05}, 
considering current-based interactions (CUBA model) and/or conductance-based interactions (COBA model). 
In our context, current based interactions correspond to gap junctions, while conductance-based interactions correspond to synaptic junctions. 
It was useless to reproduce the original benchmarks in \cite{brette-rudolph-etal:07} or \cite{vogels-abbott:05}, but 
interesting to experiment if we can explore the network dynamics with the improved model proposed here, using the method proposed in~\ref{gif-reduction}, 
and the parameters reviewed in appendix~\ref{gif-normalization}, thus beyond CUBA/COBA models.

\begin{figure}[htb]
\centerline{\begin{tabular}{cc}\parbox[u]{4cm}{\begin{tabular}{c} 
 \includegraphics[width=4cm,height=4cm]{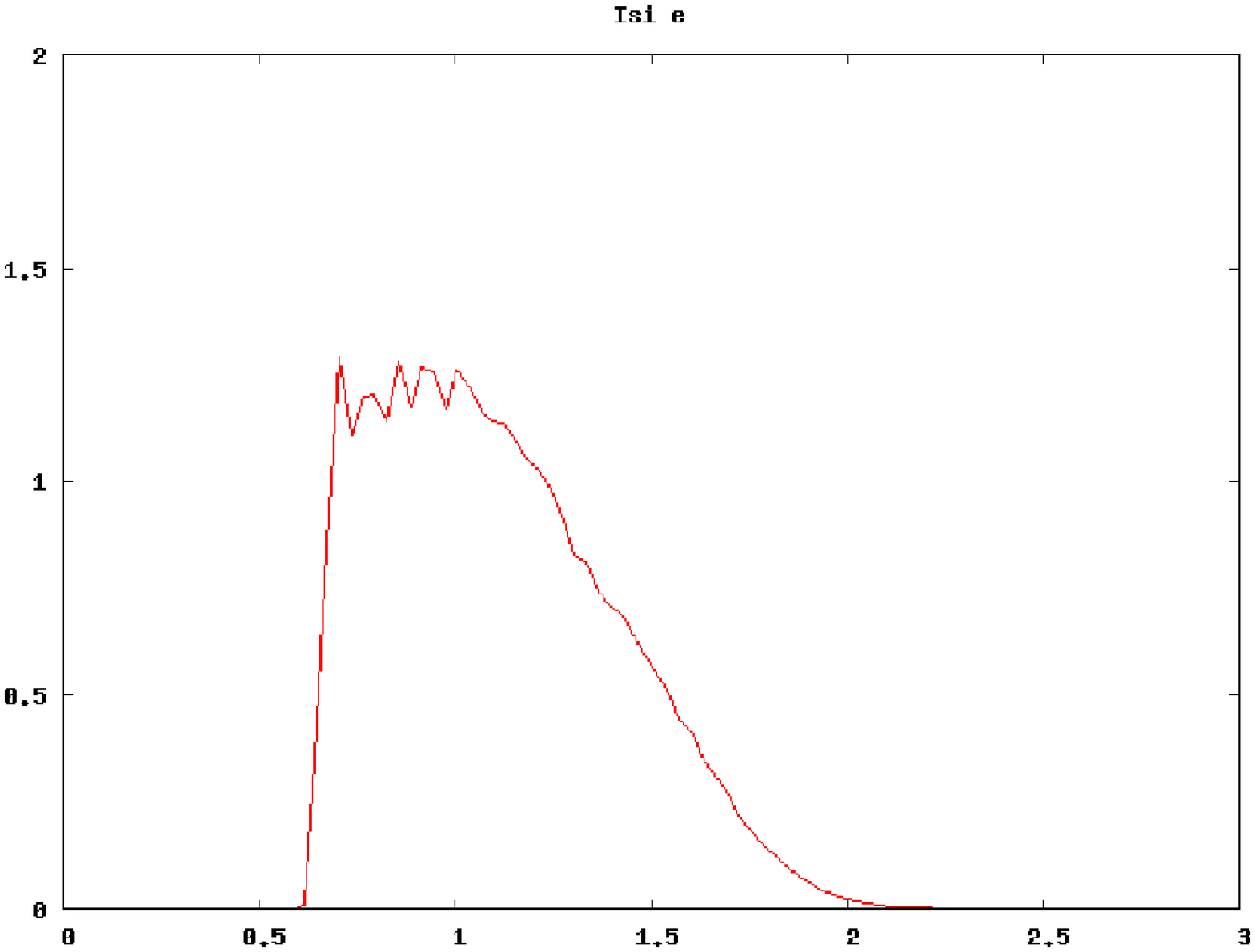} \\ \includegraphics[width=4cm,height=4cm]{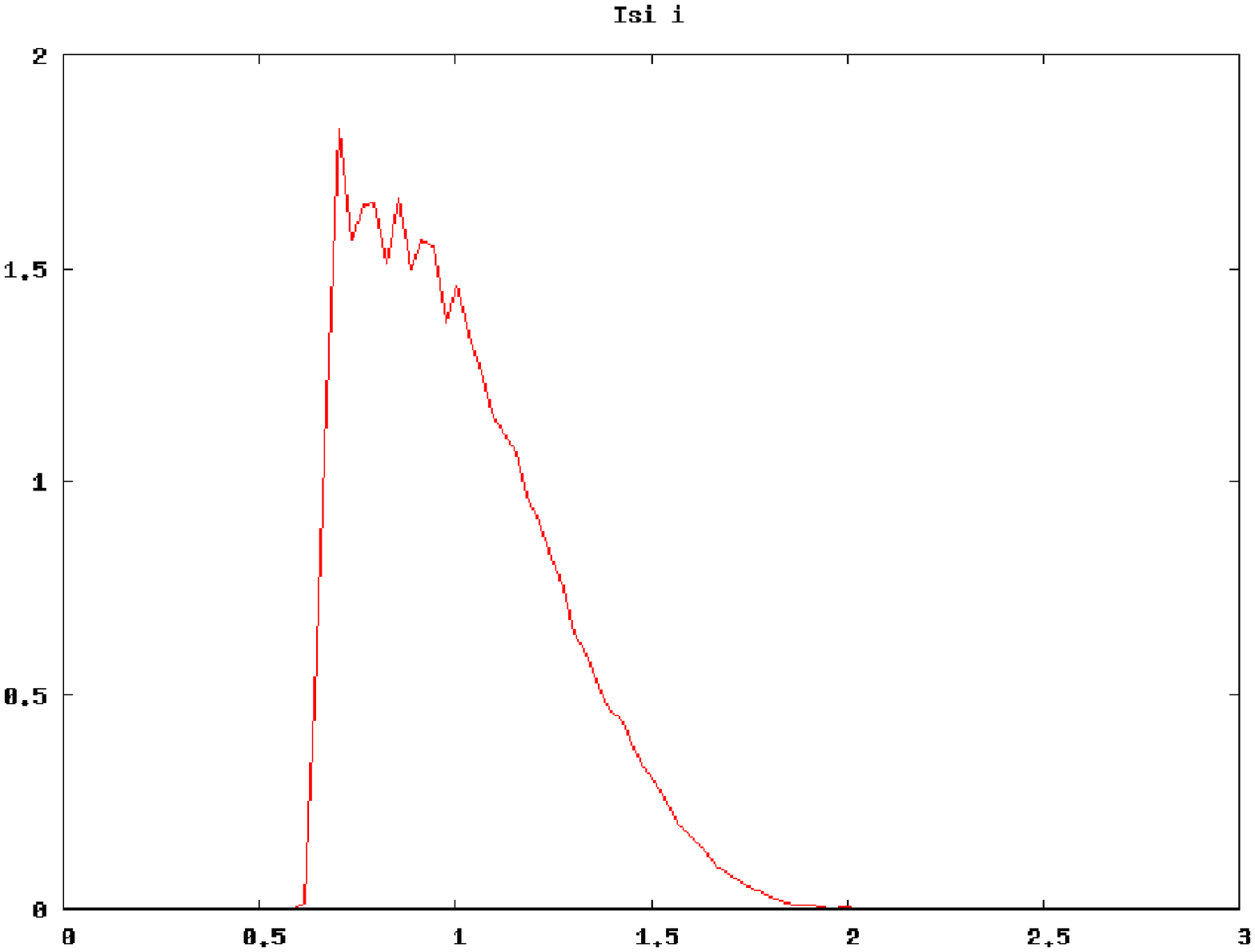} \\
 \end{tabular}} & \parbox[u]{12cm}{\includegraphics[width=12cm,height=8cm]{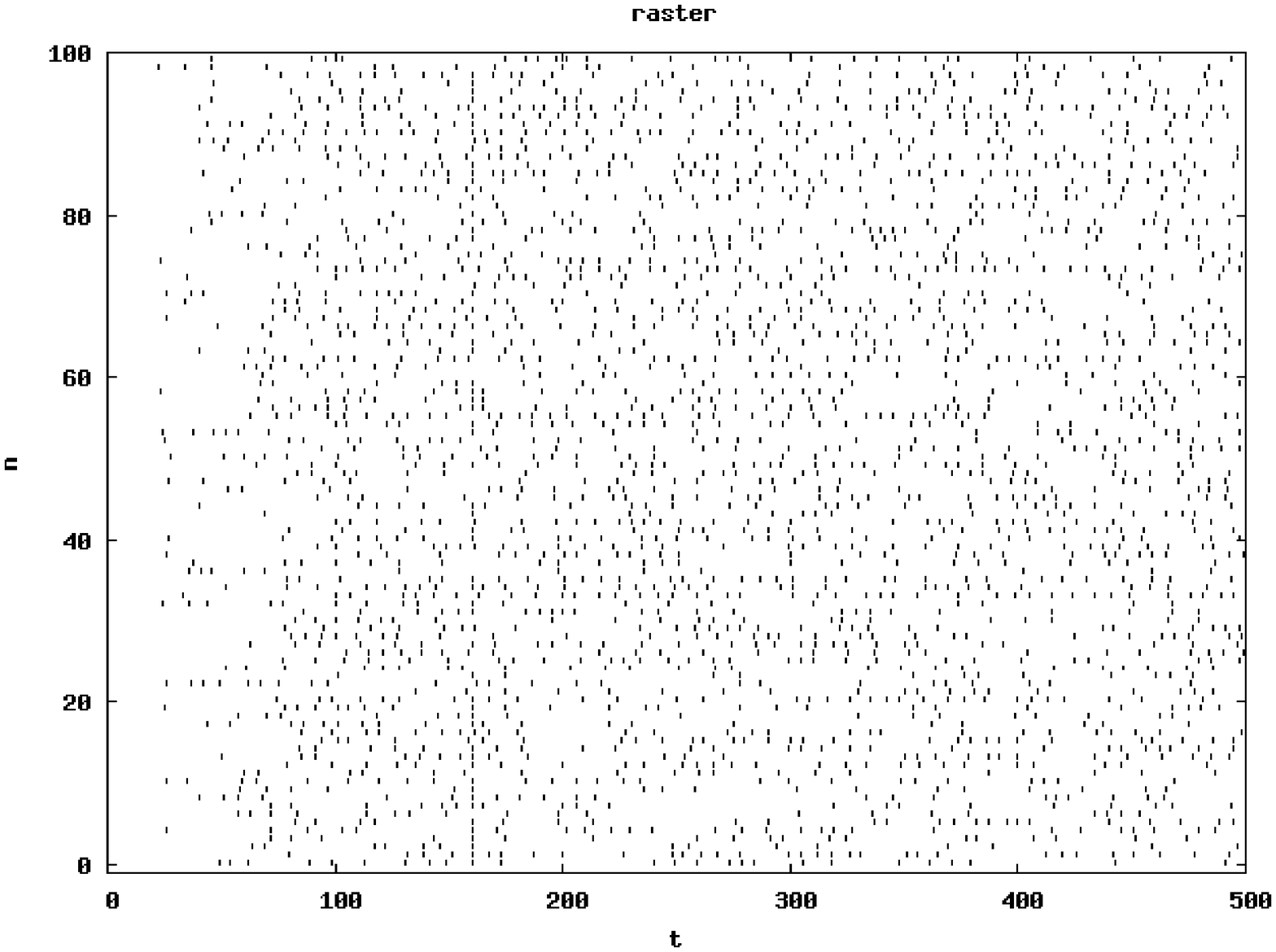}} \end{tabular}
}
\caption{\label{histo} Inter-spike interval histogram for excitatory neurons (top-left) and inhibitory neurons (down-left) with the corresponding raster head (right).
The abscissa is the decimal of the histogram log of the interval and the ordinate the inter-spike observed probability.}
\end{figure}

One result is shown in Figs.~\ref{histo}. Results are coherent with what is discussed in details in \cite{vogels-abbott:05},
and in particular close to what has been reviewed in \cite{vogels-etal:05}. This is clearly a preliminary test and the influence, on the network dynamics,
of thus alternate model is out of the scope of the present work, and a perspective for a further study.

\section{Discussion}

Taking global temporal constraints into account, it has been possible to better understand, 
at the simulation levels to which extends spiking mechanisms are bounded and simplified. 
At this simulation level, the challenge is to generate spike-trains corresponding 
to what is observed in biology or to what is required for computational processing, without the necessity to {\em precisely reproduce the internal neuron state}. 
This is a very important simplification when the goal is to switch from the neural scale to the network one.

The proposed mechanism is a complement of existing simulation tools \cite{brette-rudolph-etal:07} in the following quantitative and qualitative senses:

\subsubsection*{Quantitative complementarity}

As a software module, it has been designed to be as fast as possible. 

The cost for this choice is that a programmatic interface is required, while in order to be available on any platform with the fastest performances,
a {\tt C/C++} implementation is required (interfaces to other programming languages being available).
The fact that it is also a ``small kernel'' allows us to target embedded applications: since computing with spikes is now a mature methodology, 
a tool to run such algorithms on various platforms (e.g., in robotics) or embedded systems (e.g., intelligent reactive devices) was required. 

This has been possible here, without any loss in precision, the underlying data-structures being strongly simplified,
but with another drawback: the network dynamics is constrained since spiking units must verify temporal constraints.

A step further, the use of models at the edge of the state of the art, such adaptive non-linear gIF networks, or SRM networks is made possible in an event-based framework,
thus with expected better performances.

Regarding gIF networks, \cite{brette:07} has proposed a pure event-based method taking step-wise synapses with exponential decays into account,
The same level of modeling has been proposed by \cite{morrison-mehring-etal:05}, mixed with clock-based mechanisms,
while \cite{rudolph-destexhe:06} have investigated how to take synaptic alpha profiles into account. 
The proposed methods are based on sophisticated analytical derivations with the risk of having a rather huge number of operations to perform at each step.
As a complementary variant of these methods, we propose here to introduce another degree of freedom, using iterative lower-bound estimations of 
the next spike time. This heuristic applied to gIF neurons seems to converge quickly in practice.

Regarding SRM networks, we have generalized the simple idea to use piece-wise linear profiles, approximating the original exponential profiles roughly 
(replacing the exponential curve by a line-segment) or at any level of precision (approximating the exponential curve by any number of line-segments).
The precision/performance trade-off is thus adjustable at will.

The reason to consider gIF and SRM neuron simulation here is that they correspond, up to our best knowledge, to the most interesting punctual neuron models actually used in 
biologically plausible neural network simulation, in the deterministic case.

\subsubsection*{Qualitative complementarity}

Two key points allow us to performs new simulations with this tool: 

 Event-based and clock-based mechanisms can be easily mixed here {\em in an event-based simulation mechanism}, whereas other implementations mix
clock-based and event-based mechanisms in a clock-based simulator (e.g., in \cite{morrison-mehring-etal:05}), or use spike-time interpolation mechanisms
in order to better approximate event-based mechanisms in such clock-based environment. Using an event-based simulator to simulate a clock is obvious, 
but usually stupid, because the event-based mechanism usually generates an heavy overhead, thus making the clock-based part of the simulation intractable.
This is not the case here, since we use this minimal data-structure and have been able to see that the overhead is less than one micro-second on a standard laptop.
It is thus appears a good design choice to use an event-based simulation mechanism to mixed both strategies. 

 The second key point is that, we have proposed a way to {\em consider adaptive non-linear gIF networks in an event-based framework}. 
It is easy to get convinced that 2D integrate
and fire neurons differential equations with non-linear ionic currents (e.g. exponential, quartic or quadratic \cite{touboul:07}) do not have closed-form solutions
(unless in very special cases). Therefore, the next spike time is not calculable, except numerically, and the exact event-based implementation is not possible.
Alternatives strategies have been proposed such as simulations with constant voltage steps \cite{zheng-etal:08} allowing to implement quadratic 1D (thus without adaptive currents)
gIF networks in a modified event-base framework. In order to get rid of these limitations, one proposal developed here is to consider adaptive currents which depends only 
on the previous spiking time neuron state and non-linear ionic currents updated only at the each incoming spike occurrence. With these additional approximations, 
the event-based strategy can be used with such complex models. This is a complementary heuristic with respect to existing choices.

 We notice that the present study only considers deterministic models, while the simulation of stochastic networks is also a key issue. Hopefully, event-based implementations
of network of neurons with stochastic dynamics is a topic already investigated, both at the computer implementation level \cite{rochel:04} and modeling level
\cite{reutimann-etal:03}. In the latter case, authors propose to reduce the multiple stochastic neuronal input activity to a dedicated stochastic input current, and
investigate this choice of modeling in an event-based framework, making explicit very good performances. This method seems to be easily implementable in our present kernel, 
though this is out of the scope of the present work.

\subsubsection*{Conclusion}

At a practical level, event-based simulation of spiking networks has been made available, using the simplest possible programmatic interface, 
as detailed previously. The kernel usage has been carefully studied, following the analysis proposed in \cite{brette-rudolph-etal:07} and detailed in Table~\ref{features}

The present implementation thus offers a complementary alternative with respect to existing methods, and allows us to enrich the present spiking network simulation 
capabilities.

\appendix

\section{Appendix: About gIF model normalization} \label{gif-normalization}

Let us review how to derive an equation of the form of~(\ref{gIFe}). 
We follow \cite{izhikevich:04,brette-gerstner:05,rudolph-destexhe:06} in this section. We consider here a voltage dynamics of the form:
\[ 
\frac{d V}{dt} + I^{leak} + I^{syn} + I^{gap} = I^{ext} + I^{adp} + I^{ion}
\]
thus with leak, synaptic, gap-junction, external currents discussed in this section.

\subsubsection*{Membrane voltage range and passive properties} 

The membrane potential, outside spiking events, verifies: \\ \centerline{$V(t) \in [ V_{\mbox{reset}}, V_{\mbox{threshold}}]$}\\
 with typically $V_{\mbox{reset}} \simeq E_L \simeq -80mV$ and a threshold value $V_{\mbox{threshold}} \simeq -50mV \pm 10mV$.
\iftrue
When the threshold is reached an action potential of about 1-2 ms is issued and followed by refractory period of 2-4 ms (more precisely,
   an absolute refractory period of 1-2 ms without any possibility of another spike occurrence followed by a relative refraction to other firing). Voltage peaks are
   at about $40 mV$ and voltage undershoots about $-90 mV$. The threshold is in fact not sharply defined.
\else
 as illustrated in Fig.~\ref{spike-potential}.

 \begin{figure}[htb] 
   \centerline{\includegraphics[width=0.8\textwidth,height=4cm]{spike-potential}}
   \caption{\label{spike-potential} A schematic view of the membrane potential. Between two action potentials, i.e. spikes, the potential is between its ``reset'' and
   ``threshold'' value. When the threshold is reached an action potential of about 1-2 ms is issued and followed by refractory period of 2-4 ms (more precisely,
   an absolute refractory period of 1-2 ms without any possibility of another spike occurrence followed by a relative refraction to other firing). Voltage peaks are
   at about $40 mV$ and voltage undershoots about $-90 mV$. The threshold is in fact not sharply defined, as discussed in the text.}
 \end{figure}
\fi

 The reset value is typically fixed, whereas the firing threshold is inversely related to the rate of rise of the action potential upstroke \cite{azouz-gray:00}.
 Here it is taken as constant.
 This adaptive threshold diverging mechanism can be represented by a non-linear ionic current \cite{izhikevich:04,brette-gerstner:05}, 
 as discussed in section~\ref{gif-reduction}.

From now on, we renormalize each voltage between $[0, 1]$ writing: \begin{equation} \label{renormalization}
 v = \frac{V -  V_{\mbox{reset}}}{V_{\mbox{threshold}} - V_{\mbox{reset}}}
\end{equation}

The membrane leak time constant $\tau_L \simeq 20ms$ is defined for a reversal potential $E_L \simeq -80 \, mV$, as made explicit in~(\ref{gIFe}).

The membrane capacity $C = S \, C_L \simeq 300 pF$, where $C_L \simeq 1 \, \mu F cm^{-2}$ and the membrane area $S \simeq 38.013 \, \mu m^2$, 
  is integrated in the membrane time constant $\tau_L = C_L / G_L$ where $G_L \simeq 0.0452 \, mS cm^{-2}$ is the membrane passive conductance.

From now on, we renormalize each current and conductance divided by the membrane capacity. Normalized conductance units are $s^{-1}$ and normalized current units $V/s$.

\subsubsection*{Synaptic currents}

In conductance based model the occurrence of a post synaptic potential on a synapse results in a change of the conductance of the neuron. 
Consequently, it generates a current of the form:
\[
I^{syn}(V, \tom_t,t) = \sum_j G_{j}^+(t, \tom_t) \, \left[ V(t) - E_+ \right] + \sum_j G_{j}^-(t, \tom_t) \, \left[ V(t) - E_- \right],
\]
for excitatory $+$ and inhibitory $-$ synapses, where conductances are positive and depend on previous spike-times $\tom_t$.

\begin{figure}[htb] 
   \centerline{\includegraphics[width=0.8\textwidth,height=3cm]{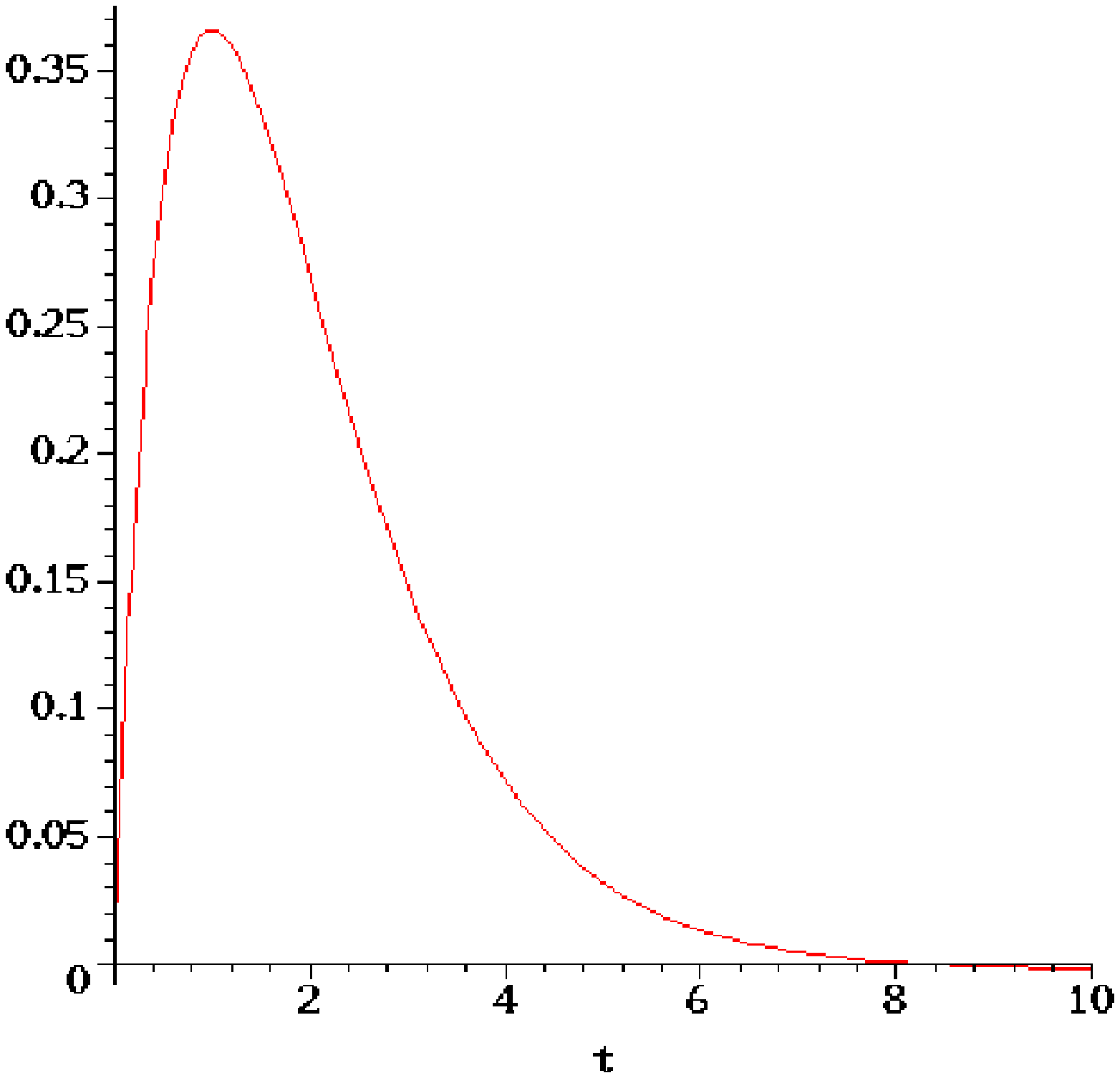}}
   \caption{\label{alpha-curve} The ``alpha'' profile $\alpha(t) = H(t) \, \frac{t}{\tau} \, e^{-\frac{t}{\tau}}$ plotted here for $\tau=1$. It is maximal at $t=\tau$ 
   with $\alpha(\tau)=1/e$, the slope at the origin is $1/\tau$ and its integral value $\int_{0}^{+\infty} \alpha(s) \, ds = \tau$ 
   since $(\int \alpha)(t) = (\tau - t) \, e^{-\frac{t}{\tau}} + k$. This profile is concave for $t \in ]0, 2\, \tau[$ and convex for $t \in ] 2 \, \tau, +\infty[$, 
   while $\alpha(2 \, \tau)=2/e^2$ at the inflexion point.}
 \end{figure}

\begin{figure}[htb] 
   \centerline{\includegraphics[width=0.8\textwidth,height=3cm]{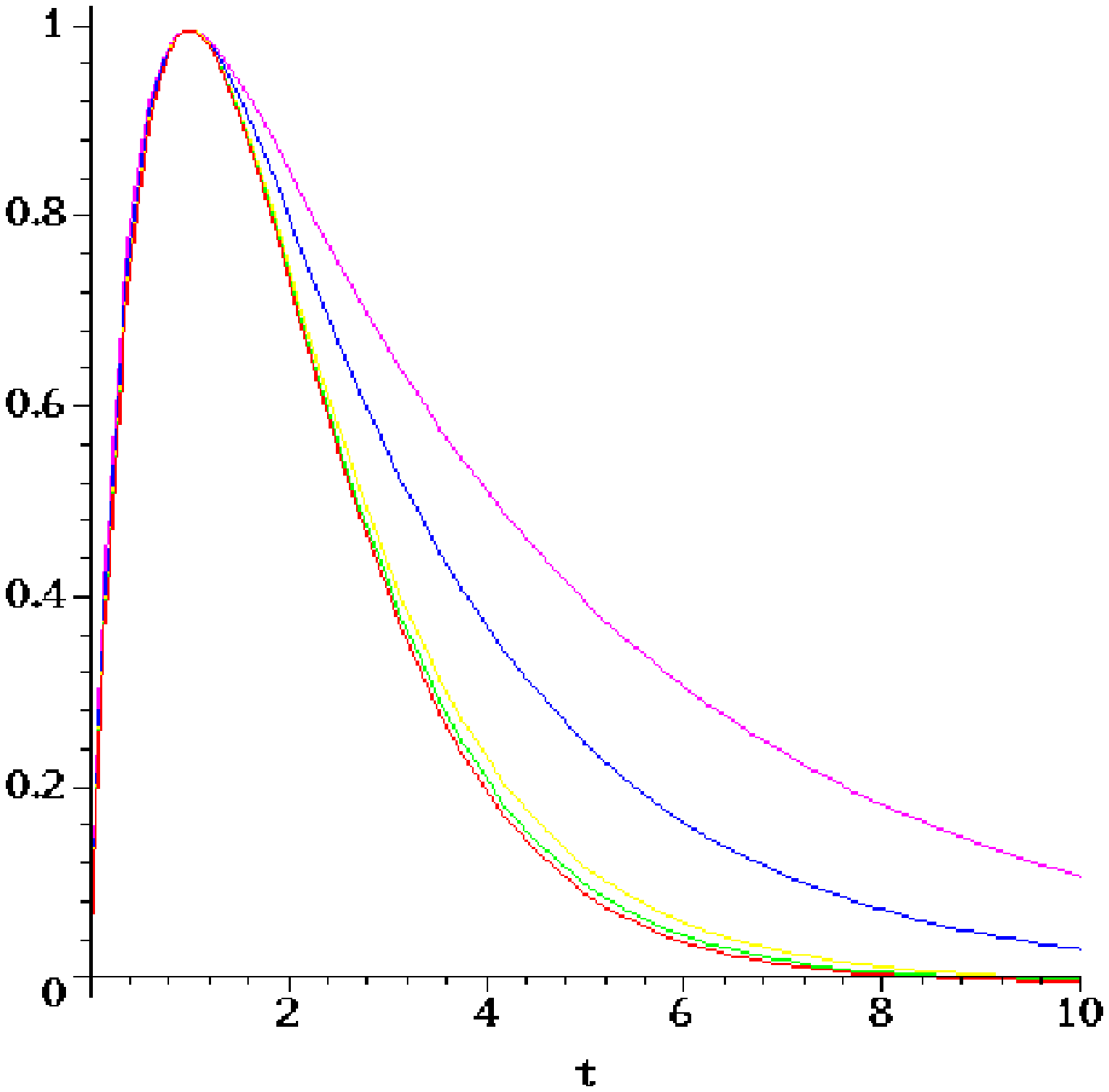}}
   \caption{\label{beta-curve} The two-states kinetic or ``beta'' profile $\beta(t) = H(t) \, \frac{1}{\kappa - 1} \, (e^{-\frac{t}{\tau}} - e^{-\kappa\,\frac{t}{\tau}})$
    is plotted with a normalized magnitude for the same $\tau = 1$ as the ``alpha'' profile
    but different $\kappa = 1.1,1.5,2,5,10$ showing the effect of this additional degree of freedom, while $\mbox{lim}_{\kappa \rightarrow 1} \beta(t) = \alpha(t)$. 
   The slope at the origin and the profile maximum can be adjusted independently with ``beta'' profiles.
   It is maximal at $t_\bullet=\tau\,ln(\kappa) / (\kappa-1)$ the slope at the origin is $1/\tau$ and its integral value $\tau/\kappa$. 
   The profile is concave for $t \in ]0, 2\, t_\bullet[$ and convex for $t \in ] 2 \, t_\bullet, +\infty[$.}
 \end{figure}

In the absence of spike, the synaptic conductance vanishes \cite{koch:99b}, and spikes are considered having an additive effect:
\[
G_{j}^\pm(t, \tom_t) = \bar{G}^\pm \, \sum_n r^\pm(t - t_j^n)
\]
while the conductance time-course $r^\pm(t - t_j^n)$ is usually modeled as an ``exponential'', ``alpha'' (see Fig.~\ref{alpha-curve}) or 
two-states kinetic (see Fig.~\ref{beta-curve}) profile, where $H$ is the Heaviside function (related to causality).

Note, that the conductances may depend on the \textit{whole past history} of the network, via $\tom_t$. 

The ``exponential'' profile ($r(t) = H(t) \, e^{-\frac{t}{\tau}}$) introduces a potentially spurious discontinuity at the origin. 
The ``beta'' profile is closer than the ''alpha'' profile to what is obtained from a bio-chemical model of a synapse.
However, it is not clear whether the introduction of this additional degree of freedom is significant here.
Anyway, any of these can be used for simulation with the proposed method, since their properties correspond to what is stated in Fig.~\ref{response}.

There are typically, in real neural networks,  $10^4$ excitatory and about $2 \,10^3$ inhibitory synapses. The corresponding reversal potential 
are $E_+ \simeq 0 \, mV$ and  $E_- \simeq -75 \, mV$, usually related to AMPA and GABA receptors. 
In average: $\bar{G}^+_{j} \simeq 0.66 \, nS, \tau^+ \simeq 2 \, ms$ and $\bar{G}^-_{j} \simeq 0.63 \, nS, \tau^- \simeq 10 \, ms$,
for excitatory and inhibitory synapses respectively, thus about $570 ms^{-1}$, $600 ms^{-1}$ in normalized units, respectively.
The coefficients $\bar{G}^\pm$ give a measure of the synaptic strength (unit of charge) and vary from one synapse to another and are also subject to adaptation.

This framework affords straightforward extensions involving synaptic plasticity (e.g. STDP, adjusting the synaptic strength), not discussed here.

\subsubsection*{Gap junctions}

It has been recently shown, that many local inter-neuronal connections in the cortex are realized though electrical gap junctions \cite{galarreta-hestin:01}, 
this being predominant between cells of the same sub-population \cite{amitai-etal:02}. At a functional level they seem to have an important influence
on the synchronicity between the neuron spikes \cite{lewis-rinzel:03}. Such junctions are also important in the retina \cite{wohrer-kornprobst:08}.

The electrotonic effect of both the sub-threshold and supra-threshold portion of the membrane potential $V_{j}(t)$ of the pre-junction neuron of index $j$ seems 
an important component of the electrical coupling. This writes \cite{wohrer-kornprobst:08,lewis-rinzel:03}:
\\ \centerline{$I^{gap}(V, t) = \sum_j G^*_{j} \, \left[ \left[ V_j(t) - V(t) \right] + E_\bullet \, \sum_{n} r(t- t_j^n) \right]$} \\
where $G^*_{j}$ is the electrical coupling conductance, the term $V_j(t) - V(t)$ accounts for the sub-threshold electrical influence 
while and $E_\bullet$ parametrizes the spike supra-threshold voltage influence. 

Regarding the supra-threshold influence, a value $E_\bullet \simeq 80 mV$ corresponds to the usual spike voltage magnitude
of the spiking threshold, while $\tau_\bullet \simeq 1 ms$ corresponds to the spike raise time. Here $r()$ profile accounts for the action potential itself, 
slightly filtered by the biological media, while the gap junction intrinsic delay is of about $10 \mu s$. 
These choices seem reasonable with respect to biological data \cite{galarreta-hestin:01,lewis-rinzel:03}

A step further, we propose to neglect the sub-threshold term for three main reasons. 
On one hand, obviously the supra-threshold mechanism has a higher magnitude than the sub-threshold mechanism, since related to action potentials.
Furthermore, because of the media diffusion, slower mechanisms are smoothed by the diffusion, whereas faster mechanisms better propagate.
On the other hand, this electrical influence remains local (quadratic decrease with the distance) and is predominant between cells of the same sub-population 
which are either synchronized or 
have a similar behavior, as a consequence $|V_j(t) - V(t)|$ remains small for cells with non-negligible electrical coupling. Furthermore, 
a careful analysis of such electrical coupling \cite{lewis-rinzel:03} clearly shows that the sub-threshold part of the contribution has not an antagonist effect on the 
neuron synchrony, i.e., it could be omitted without changing qualitatively the gap junction function.

As a conclusion, we are able to take gap junction into account with a minimal increase of complexity, since we obtain a form similar to synaptic currents, 
using very different parameters

\subsubsection*{External currents} 

Direct input (or external) current is often related to electro-physiological clamps. 
At another level of representation, the average activity of the neuron can be modeled by a constant or random input current.
In both cases, the way the proposed simulation methods is proposed requires to assume such external current to be taken as constant between two spikes, 
i.e. having temporal variations small enough to be neglected.
If not, it is easy to associate to the external current an event unit which fires at each new current value, in order the neuron to take this new value into account.

\bibliographystyle{apacite} 

{\scriptsize \bibliography{../../../Latex/string,../../../Latex/odyssee,../../../Latex/biblio}}

{\small {\bf Acknowledgment:} Partially supported by the ANR MAPS \& the MACCAC ARC projects. 
We are especially thankful to H\'el\`ene Paugam-Moisy for precious advices and inputs regarding this piece of work.}

\clearpage 
\begin{table}[htb] 
\centerline{\small \begin{tabular}{|ll|l|}
\hline \multicolumn{3}{|c|}{Features} \\ \hline
Clock-based      &: can it simulate clock-based strategies ?                    & yes \\
                 &: in this case, does it use extrapolation for spike times ?   & useless$^1$ \\
Event-based      &: can it simulate event-based strategies ?                    & yes \\
                 &: in this case, is the integration scheme exact ?             & yes$^2$ \\
Parallelism      &: does it support parallel processing ?                       & no$^7$ \\
Graphics         &: does it have a graphical interface ?                        & no, but a programmatic interface \\
Simple analysis  &: is it possible to perform simple analysis ?                 & yes with visualization \\
Complex analysis &: can more complex analysis be done ?                         & it can$^3$ \\
Interface        &: is interface to outside signals possible ?                  & indeed$^4$ \\
                 &: is it interoperable with other simulators ?                 & yes$^4$ \\
Save option      &: can simulation be halted / resumed ?                        & yes \\
Neuron models    &: can it simulate HH models ?                                 & it can$^5$ \\
                 &: can it simulate leaky IF models ?                           & yes \\
                 &: can it simulate multivariate IF models ?                    & it can$^5$ \\ 
                 &: can it simulate conductance-based synaptic interactions ?   & yes \\
                 &: can it simulate short-term plasticity ?                     & it can$^5$ \\ 
                 &: can it simulate long-term plasticity ?                      & it can$^5$ \\ 
                 &: can it simulate compartmental models with dendrites ?       & no \\
\hline \multicolumn{3}{|c|}{Usage} \\ \hline
Development      &: is it currently developed ?                                 & yes, still $\alpha$-version \\
                 &: how may developers yet ?                                    & half-time researcher + students \\
Support          &: is it supported                                             & yes \\
                 &: what kind of support                                        & email + phone \\
                 &: are they user cooperative tools ?                           & not yet, tools available$^6$ \\
Manual           &: are there tutorials and reference material available ?      & yes \\
                 &: are there published books on the simulator ?                & no (useless) \\
                 &: is there a list of publications of articles that used it ?  & yes \\
Import/export    &: is standard (XML) specification import/export available ?   & it can$^3$ \\
Web site         &: is there a web site where all can be found ?                & {\tt http://enas.gforge.inria.fr} \\
Source code      &: are there codes available on the web ?                      & yes \\
Operating system &: does it run under Linux                                     & yes, tested \\
                 &: does it run under Max-OS X                                  & yes, tested \\
                 &: does it run under Windows                                   & likely (untested) \\
Interoperability &: using which language can it be used ?                       & C/C++, Python, Java, Php \\
                 &: can it be used from other platforms ?                       & yes$^8$ \\
\hline \end{tabular}}
Notes: {\scriptsize
\\ $1$: Since clock/event-based mechanisms can be mixed in a event-based simulation, spike times extrapolation is no more to be used, but exact event-time instead.
\\ $2$: Exact integration scheme is to be used when allowed by the model, lower-bound spike time evaluation is a new alternative proposed here when the former is not possible.
\\ $3$: More complex analysis and XML specification import/export is indeed possible, using this kernel within the PyNN environment. The goal was to {\em only} develop here,
what was not available elsewhere. The API has been carefully designed for this purpose.
\\ $4$: The interface capability is a key feature of this middle-ware implementation, including in real time applications using spike computations.
\\ $5$: Plasticity and other existing models, not discussed here, can be implemented with this middle-ware, and STDP is already considered at that time. The real goal
is however not to implement ``all'' models, other simulators do that better, but to propose also alternatives to existing models, as discussed in this paper.
\\ $6$: The present development is installed on a forge, thus has all forum/bug-tracking/user-resquest-ticketting, etc.. available.
\\ $7$: Though parallelism is not available yet, and the issue not addressed here, the network simulation kernel has been designed and implemented in order to be able to connect 
to other kernels, via input/output events. It is thus a feasible extension to run several kernels in parallel, with the drawback that the slower kernel is going to
drive other kernels local times.
\\ $8$: Links with these external platforms such as PyNN (thus NEURON, MvaSpike, ..), NeuroConstruct are made available by the multi-language operability, 
while Scilab and Matlab usage is documented.
}
\caption{\label{features} Summary of the main features of the implemented event-based simulation kernel, using the criteria proposed to compare existing simulators,
see text for details. The required features have been set by the authors group of (Brette et al., 2007) and applied to almost all existing simulators at this date.}
\end{table}

\end{document}